\documentclass[useAMS,usenatbib]{mn2e}

\usepackage{graphics}
\usepackage{graphicx}
\usepackage{amssymb}

\title[Triangulating Radiation]{Triangulating Radiation:\\
Radiative Transfer on Unstructured Grids}

\author[J. Ritzerveld, V. Icke, and E.-J. Rijkhorst]{J. Ritzerveld$^{1}$\thanks{E-mail:
ritzerveld@strw.leidenuniv.nl}, V. Icke$^{1}$ and E.-J. Rijkhorst$^{1}$\\
$^{1}$Sterrewacht Leiden, P.O. Box 9513, 2300 RA Leiden, The Netherlands}

\begin{document}

\date{Accepted. Received}

\pagerange{\pageref{firstpage}--\pageref{lastpage}} \pubyear{2004}

\maketitle

\label{firstpage}

\begin{abstract}
We present a new numerical approach that is able to solve the multi-dimensional
radiative transfer equations in all opacity regimes on a Lagrangian,
unstructured network of characteristics based on a stochastic point process. 
Our method reverses the limiting procedure used to derive the transfer equations,
by going back to the original Markov process. Thus, we reduce this
highly complex system of coupled differential equations to a simple
one-dimensional random walk on a graph, which is shown to be
computationally very efficient. Specifically, we use a Delaunay graph,
which makes it possible to combine our scheme with a new smoothed
particle hydrodynamics (SPH) variant proposed by \citet{Pelupessy}. 
We show that the results of applying a two-dimensional implementation of our
method with various suitable test cases agree with the analytical
results, and we point out the advantages of using our method with
inhomogeneous point distributions, showing examples in the progress.
Hereafter, we present a supplement to our method, which can be useful
in cases where the medium is optically very thin, and we conclude 
by stating some anticipated properties of this method
in three dimensions, and announce future extensions.
\end{abstract}

\begin{keywords}
Radiative transfer -- Methods: numerical -- Hydrodynamics
\end{keywords}

\section{Introduction}

The formation of cosmic structures, such as galaxies and stars, is
almost certainly dominated by an intricate interplay between (magneto)hydrodynamics,
gravity, and radiative transfer, on a cosmological background that
sets the initial and boundary conditions. Of these, the cosmology
is assumed to be given, while the computation of gravitational potentials
is rather well understood \citep[e.g.][]{Greengard}. Hydrodynamics must
be three-dimensional for this purpose, and 3D hydro is beginning to
enter its springtime: adaptive-mesh refinement (AMR) and related methods
are beginning to produce results \citep[see e.g.][for a review]{LeVeque}.
However, radiative transfer techniques that combine true three-dimensionality
with reasonable spectral resolution are, by comparison, the most primitive
of the methods needed for realistic simulation of structure formation.
Yet it seems essential that the physics of radiation be built in,
because the energy budget of nascent structures is heavily influenced,
indeed sometimes dominated, by radiative effects. Serious models must
be three-dimensional, and spectral coverage must at least be good
enough to cover hydrogen ionisation, recombination and photodissociation.
This makes solving the radiation part of the physical problem seven-dimensional:
a daunting computational task.

Usually, a galaxy is represented by a finite point set, with a size
of the order of \( 10^{6} \), in which each point represents a fixed
mass fraction of the galaxy. Because these points represent thousands
of solar masses or more, they are not actually point-like. Thus, one
is immediately confronted with the problem that the laws of motion
are differential equations that represent a continuum, which cannot
be uniquely defined on a discrete point set. In order to circumvent
this problem, one can convolve the set with a smoothing function so
that a continuous field is obtained. In this way, gradients and other
derivatives are properly defined. The scheme which combines this smoothing
trick with the implementation of the equations which govern the dynamical
behaviour of fluids, is called \emph{Smoothed Particle Hydrodynamics}
\citep[SPH;][]{Lucy}. It has been well established that the use of SPH,
under certain restrictions, can be very fruitful for doing astrophysical
hydrodynamical calculations \citep[for a review, see][]{Monaghan}.

Next, the interaction between radiation and matter must be included.
The radiative transfer equations, which describe this interaction
macroscopically, are a system of non-local, coupled differential equations
and are extremely difficult to solve analytically and numerically
\citep[e.g.][]{Rutten}. Recently, \citet{Pelupessy} stated that
it would be advantageous in several ways to avoid the use of a smoothing
function when using SPH, and instead use the Delaunay tessellation
of the point set to create a continuous field. This method is called
the \emph{Delaunay Tessellation Field Estimator} \citep[DTFE;][]{Schaap}. 
Accordingly, if we could find a scheme that
is able to solve the radiative transfer equations on a Delaunay grid,
we could combine the two, and thereby introduce radiative transfer
in a natural and possibly economical way into particle-based methods
such as SPH.

The aim of this paper is, to present a new numerical method that is
able to solve the radiative transfer equations by means of a Markov
process on networks of characteristics, such as a Delaunay graph. We adopt the
Lagrangian treatment of the SPH-scheme and let the point process represent
the underlying mass distribution, by which the method will be able
to solve the equations in all opacity regimes. The method is extremely
fast, conceptually very simple, and because of its generic setup it
is applicable in spaces of any dimension.

First, we discuss the extant numerical schemes for solving
the transfer equations (in three dimensions), emphasising their advantages
and disadvantages, and why these are not sufficient for the needs
at hand. Second, we present our new method, after which we show the
results of using our method with several two-dimensional test cases. Thereafter,
we point out the advantages of our method by using
it on a correlated, inhomogeneous point distribution. We finish by presenting
a version of our method that can be used when the
medium is optically very thin.

\section{Numerical radiative transfer}

For our present purposes, it suffices to summarise the quantum nature
of the interaction between radiation and matter by macroscopic parameters,
for example a scattering cross section or a mean absorption coefficient.
If we assume that the radiation relaxation time is small compared to the 
the other time scales of the considered physical system, we may use
the time-independent (equilibrium state) Boltzmann equation for photons
(in general \(d\) dimensional space),

\begin{equation}
\label{rad_transfer}
\bmath{n}\cdot \nabla I_{\nu }(\bmath{x},\bmath{n})=j_{\nu }(\bmath{x},\bmath{n})-\alpha _{\nu }(\bmath{x},\bmath{n})I_{\nu }(\bmath{x},\bmath{n}).
\end{equation}
 This equation relates the spatial gradient of the luminous intensity
\( I_{\nu }(\bmath{x},\bmath{n}) \) of photons with frequency \( \nu \in \mathbb{R}^{+} \)
travelling in the direction \( \bmath{n} \in S^{d-1} \), at the location \( \bmath{x} \in \mathbb{R}^{d}\),
to certain source terms. The right hand side of this equation lists
the source terms for \emph{emissivity} \( j_{\nu }(\bmath{x},\bmath{n}) \)
and for \emph{extinction} \( \alpha _{\nu }(\bmath{x},\bmath{n}) \),
which includes the scattering coefficient \( \alpha ^{\rmn{scat}}_{\nu }(\bmath{x},\bmath{n}) \)
and the pure absorption coefficient \( \alpha ^{\rmn{abs}}_{\nu }(\bmath{x},\bmath{n}) \).

Eq.(\ref{rad_transfer}) is the time-\emph{in}dependent radiative transfer equation, which describes
the radiative properties of systems in radiative equilibrium, and it is this
equation that our new method solves, as we will show in what follows. 

\subsection{Limiting behaviour}

A general analytical solution of this non-separable integro-differential equation of first order
does not exist, because
of its behaviour in different limiting (opacity) regimes. For example,
if we take a dominating scattering cross section, that is \( \alpha ^{\rmn{scat}}_{\nu }(\bmath{x},\bmath{n})D\gg 1 \),
with \( D \) as the thickness of a layer of medium,
a particular photon will be scattered many times, so that the angular
dependence of the original incident angle is wiped out. It can be
shown algebraically \citep[e.g.][]{Duderstadt} that in this
limit the transfer equation Eq.(\ref{rad_transfer}) can be rewritten
as a diffusion equation, the solutions of which are angle \emph{in}dependent.

On the other hand we have the limit in which the scattering cross
section is very small, that is \( \alpha ^{\rmn{scat}}_{\nu }(\bmath{x},\bmath{n})D\ll 1 \).
In this case, if also \( \alpha ^{\rmn{abs}}_{\nu }(\bmath{x},\bmath{n})D\ll 1 \),
the mean free path of a photon is very large, so that neighbouring
paths need not be correlated, and there may be a different solution
for each angle. In this case, the transfer equation can be rewritten
as an upwind hyperbolic PDE, the solutions of which are angle \emph{de}pendent.

Thus, the characteristic properties of the solution in different limits
can be quite contradictory, which shows the great difficulty in finding
an explicit general solution to the transfer equation.

\subsection{Numerical schemes}

There are numerous numerical schemes that are excellent at solving
the transfer equation in one of the opacity regimes. But in passing
from one regime into the other most schemes fall short. A solver for
the diffusion limit cannot solve the hyperbolic PDE, and vice versa.
Fortunately, some schemes have been developed which, in principle,
are able to solve the transfer equation in all opacity regimes. Most
of these can be subdivided into three main categories: those using
\emph{long characteristics}, \emph{short characteristics}, and the
\emph{Monte Carlo} methods. Because we want to point out how these
relate to our new method, and in particular how ours improves on the
extant ones, we briefly sketch their approach.

First, all of these methods work by superimposing a grid on the domain
on which the transfer equation has to be solved. The aim is to find
the intensity \( I_{\nu } \) in a number of directions for each of
the grid cells. In the Monte Carlo approach, one sends out \( N \) photon packets
from each grid cell in a certain number of random directions, and
one just keeps detailed track of its scattering, absorption and re-emission. 
These methods are very easy to implement, it allows for very complicated
spatial distribution and arbitrary scattering functions.
However, because they use statistical averaging, they
introduce statistical noise, which can only be suppressed by taking
a large value for \( N \). Because the noise reduction scales with
the square root of \( N \) only, Monte Carlo methods are computationally
very expensive.

The long characteristics method \citep[first suggested by][]{Mihalas} uses rays
(\emph{characteristics}) which connect a given grid cell to every
other relevant cell. The transfer equation is solved one-dimensionally
along these lines. This type of method has the advantage that it incorporates
the nonlocality of the transfer equation, and is thus able to solve
it accurately for arbitrary density configurations. 
A disadvantage is that the method becomes computationally very expensive, 
if one wants high angular resolution, so as to
accurately sample space at large distances from the source.
Moreover, the long characteristics usually
cover the same part of the domain many times. This introduces strong
redundancy, which makes the method time-consuming. 

A way around this redundancy problem is the short characteristics
method \citep[first proposed by][]{Kunasz}. In this case, 
one calculates the intensity in one grid cell
by connecting it with its neighbouring cells only, and solves the transfer
equation one-dimensionally along these lines. An advantage of this
method is, that it is not very redundant, but it also requires a very
clever scheme to sweep the grid, in order to be sure that the intensities
in all the neighbouring grid cells are known when they are needed. This is necessary
because the emissivities may depend on the intensities, for example
in the case of scattering. 
The physical values of the neighbouring cells
contribute via interpolations along the grid lines, which have to
be quadratic or higher order in order to accurately reproduce the diffusion
limit, which is governed by the second order diffusion equation.
The interpolation, intrinsic to the short characteristic methods,
introduces angular diffusion into the numerical solution, 
for example causing parallel laser beams to diverge in the downwind direction
\citep[e.g. see][]{Steinacker}.
\citet{Kunasz}
showed that a parabolic interpolation reduces this intrinsic numerical diffusion, thereby
obtaining a more accurate result, but not only does it make the algorithm more
complex, because it requires three upwind interpolation points, but it can also
cause unphysical under- and overshoots of the interpolated quantities near
discontinuities, possibly resulting in values for \( I_{\nu } \) which are negative.

\begin{figure}
\begin{center}
\includegraphics[width=4.1cm,clip=]{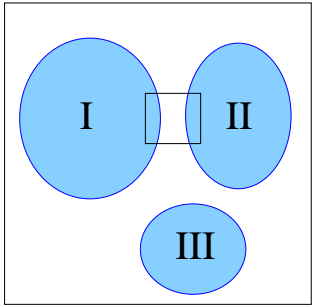} 
\end{center}
\caption{\label{Clouds}Schematic illustration of several radiating clouds 
surrounded by vacuum.}
\end{figure}

To illustrate the difficulties of these methods, we present a schematic example of several 
radiating optically thick clouds which are surrounded by vacuum (see Fig.\ref{Clouds}).
Albeit that this example is oversimplified, it is illustrative, because
one immediately sees that, if we want to calculate the radiation profile
of the emission of cloud II, we have to take into consideration the radiation
emitted by the neighbouring clouds, because this can contribute to the desired
profile via scattering and/or re-emission. Radiation emitted by cloud I will
encounter sharp gradients in the opacity, when it leaves the optically thick cloud and
streams freely into the vacuum, until it gets absorbed or
scattered by the optically thick cloud II. 

\begin{figure}
\begin{center}
\includegraphics[width=8.4cm,clip=]{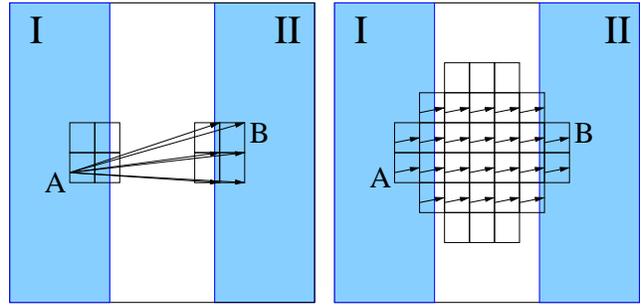} 
\end{center}
\caption{\label{CloudsZoom}Zoom in of the square in Fig.\ref{Clouds}. Illustration
of two mechanisms for calculating the effect of radiation passing through
optically very thin medium, after being emitted at a point A 
near the border of cloud I, on the medium in the neighbourhood of a point B near the
border of cloud II. Left is a schematic illustration of a long characteristic method;
right is one of a short characteristic method.}
\end{figure}

Zooming in on the square in Fig.\ref{Clouds} we obtain Fig.\ref{CloudsZoom}.
If one wants to calculate the effect of the radiation emitted at a point A
near the border of cloud I on the local re-emission properties of a point B
near the border of cloud II, one can take the long (Fig.\ref{CloudsZoom}, left)
and the short characteristics (Fig.\ref{CloudsZoom}, right) approach.
From the analytical solution we know that the radiation emitted by cloud I should 
propagate outwards as a plane wave (assuming we zoom in sufficiently for the boundary to
be a straight line), with characteristics perpendicular to the outer 
boundary of the cloud.
One thing we immediately see is that, in order to accurately simulate the plane wave, 
at least the whole boundary region has to be incorporated as a source, for example by sampling that
region by a considerable amount of point sources. Because the operation count of both methods
scales with the number of sources, an extended source like in this example will increase the
required number of operations enormously, sometimes even beyond the reach of modern
computer power. Moreover, most radiative transfer methods are designed to solve physical
problems in which there is an inherent geometrical symmetry, most frequently axi-symmetry,
or even more simple, to solve for just one point source. Of course, both these simplifications
result in severe restrictions on the type of physical configuration one would like to model.
Another complication, when using a long characteristic method (Fig.\ref{CloudsZoom}, left), 
is that the angular sampling
has to be very accurate so as to have a good sampling of space at large distances. If the
distance from cloud I to cloud II increases, the accuracy and thus the number of cycle counts
needed increases proportionally.
In the short characteristic methods (Fig.\ref{CloudsZoom}, right), we see that the upwind
interpolation scheme gives rise to numerical, unphysical diffusion in the empty region
between cloud I and II, by which the angular resolution at large distances from the source
tends to get smeared, so that detailed information is lost. This effect would be even more
dramatic in the case that the boundary region is inhomogeneous, which would give cause to
a wave front with a rich variety of multipole features.

In addition, there is a generic drawback inherent in all these classes, which we immediately see
from Fig.\ref{CloudsZoom}: they use a stiff grid which has nothing to do with the underlying physical
problem. When using those \emph{unphysical} grids, one introduces preferential directions and
superimposed scale lengths which are not related to the problem at hand. Because we need a grid
fine enough to accurately resolve the dense structure near the boundary, we enormously oversample the 
optically thin region between the clouds.

The new method, which we present in the next section, is a supplement
to the extant methods, in the sense that it is not restricted to just
one opacity regime \emph{and} that it uses a physical grid, thereby avoiding
the difficulties of using an unphysical one, as we have just described. Moreover, it does
\emph{not} scale with the number of sources, performing equally fast for extended sources
as for one point source.
It is, in a sense, a combination of all three categories of radiative transfer methods,
but we will elaborate on this comparison later on.

\section{A new method}

In his landmark paper, \citet{Chandrasekhar} showed that the migration of 
photons through a medium can be described as a Markov stochastic process.
More specifically, the migration can be described as a random walk of photons
through a medium during which they may get scattered or absorbed according
to the scattering coefficient \( \alpha ^{\rmn{scat}} \) and the absorption coefficient
\( \alpha ^{\rmn{abs}} \) of the medium. A normalised phase function, \(f(\bmath{n},\bmath{n}')\),
describes the probability of a photon scattering from direction \(\bmath{n}\) to
\(\bmath{n}'\). The free path between two consecutive events, which can either be
scattering or absorption, has an exponential distribution in the form of
\( \alpha ^{\rmn{tot}} \rmn{e}^{{-\alpha ^{tot}D}} \), which is characterised by the total
attenuation \( \alpha ^{\rmn{tot}} = \alpha ^{\rmn{scat}} + \alpha ^{\rmn{abs}} \). At one such
event, absorption takes place with a probability \( \alpha ^{\rmn{abs}} / \alpha ^{\rmn{tot}} \)
and scattering with probability \( \alpha ^{\rmn{scat}} / \alpha ^{\rmn{tot}} \). This picture
forms the basis for the Monte Carlo simulation of photon migration.
\citet{Chandrasekhar} showed that by taking a large number of steps or, equivalently,
by averaging over a large number of possible paths, one can use these microscopic statistics 
to derive macroscopic quantities,
such as the number of photons at a certain distance from the source, travelling in a certain
direction, which is of course the pivotal specific intensity.

Our new method is characterised by the approach that we sample the medium
by a finite amount of discrete event centres, in such a way that the volume average
over a certain region containing these event centres 
results in the correct macroscopic physical quantities,
such as the scattering and absorption optical depths, for the medium we try to model.
One crucial assumption is that the ensemble of scattering or absorbing particles, i.e.
event centres, is ergodic, so that the sample we choose is representative for the
whole ensemble.
The essential aspect of our new method is that we use this set of event centres
as our set of grid points, coupled with a specific choice for their interconnection, the
Delaunay/Voronoi tessellation.

\subsection{The grid}

We do not use a grid in the usual sense, 
nor do we solve a differential equation. Instead, we return to the physical origin
of the equations of radiative transfer by introducing a \emph{point
process} on which we let photons travel by a Markov process.
Thus, we use a \emph{physical} grid for radiative transfer. The placement
of the grid points (the point process) is determined by the underlying
mass distribution, 
which may adapt to the dynamical properties of
the medium. Thus, given a certain amount of available grid points,
we put most at places where the density is highest and least in low-density
areas. The exact recipe we use for placing the available points is discussed
in Subsection 3.2.2.

\begin{figure}
\begin{center}
{\includegraphics[width=4.1cm,clip=]{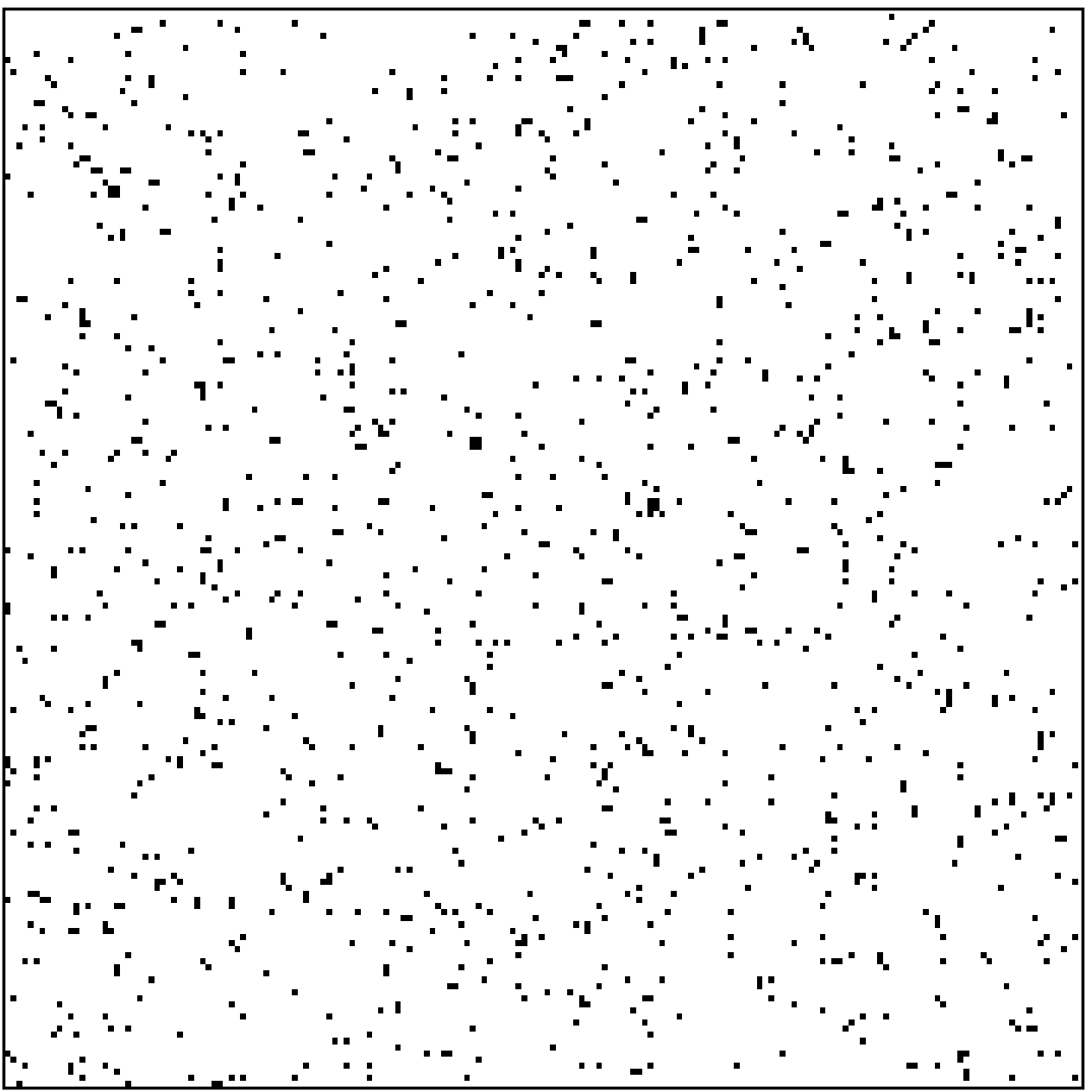}
\includegraphics[width=4.1cm,clip=]{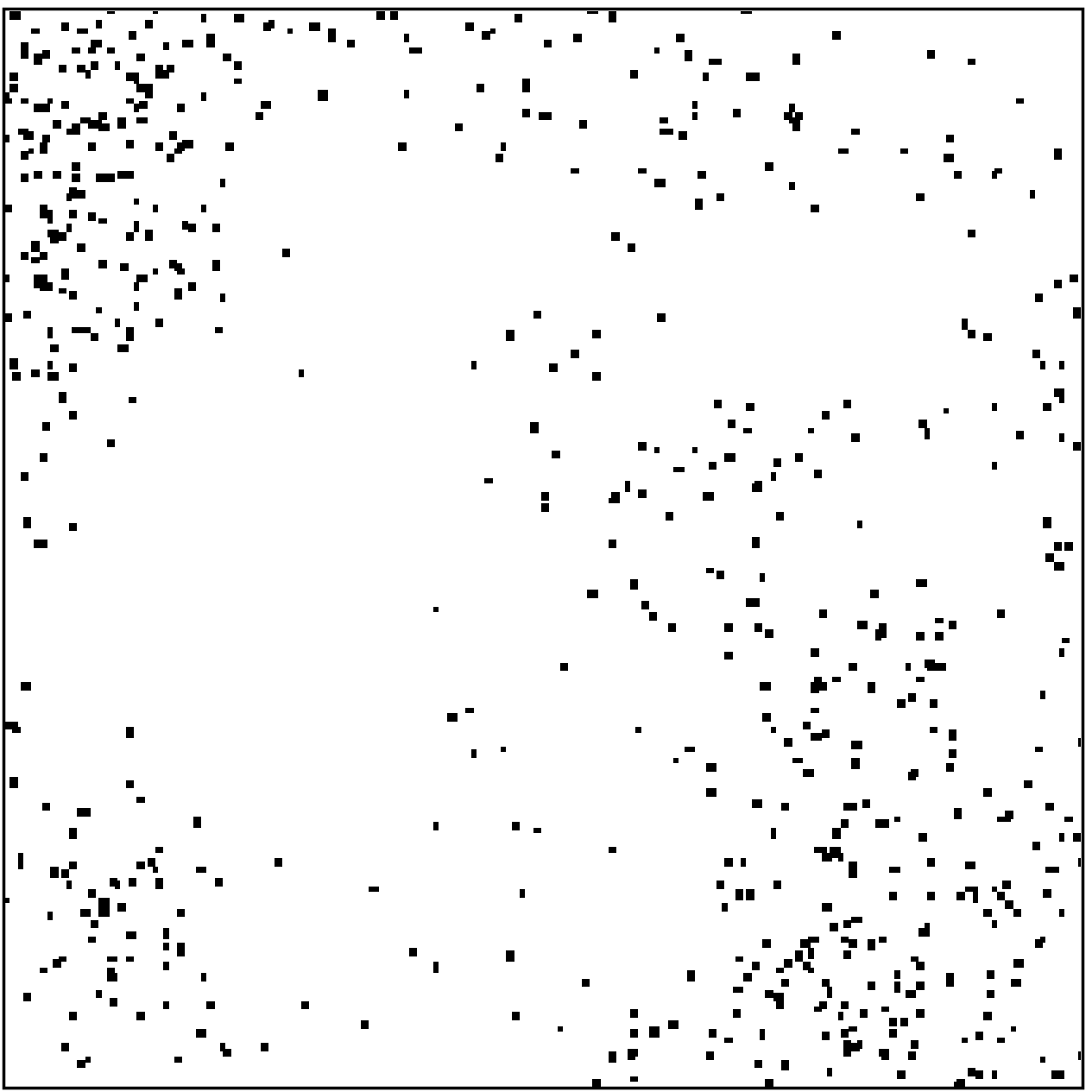}}
%{\centering \resizebox*{0.49\columnwidth}{!}{\includegraphics{Poisson.eps}} 
%\resizebox*{0.49\columnwidth}{!}{\includegraphics{Correlated.eps}} \par}

\caption{\label{Point_distr}\emph{Left}: Poisson point process representing a
homogeneous medium; \emph{Right}: Correlated point process representing a
clumpy medium.}
\end{center}
\end{figure}

A key issue of our method is that we use a stochastic point process
as a recipe for placing the points. If the underlying mass distribution
were homogeneous and isotropic, we should use a Poisson process. The
average amount of points within a certain area would be a constant
\( \rho_{\rmn{D}}  \), called the \emph{point intensity}. If the medium
distribution were inhomogeneous, we would have to use a correlated
point process, as a result of which the point distribution would show
clumpiness. Either way, we make use of a random number generator to
get the coordinates of a point. This automatically simulates a Poisson
point process (see Fig.\ref{Point_distr}, left). In the case of a
correlated point distribution, we reject some coordinates and move
on to the next in such a way that the overall distribution has a
density profile conforming to the underlying medium distribution (see Fig.\ref{Point_distr},
right). Of course, if we have the exact \(d\) dimensional density
distribution function (or a discretised \(d\) dimensional density array),
it is always possible to use Monte Carlo methods to sample
that density distribution with a finite amount of points with an accuracy
that is only limited by the number of points.

Next, we must specify a way of connecting the grid points with a network of
lines along which the photons will travel (characteristics). The nice thing
about our method is that it is not very strict in what the requirements
for such a connection scheme should be. This freedom in constructing
a network makes it possible to use our method on many different types
of grid. That said, it is of course very desirable to choose an approach
that does not suffer from the drawbacks we have just criticised, such
as the incorporation of a preferred length scale or preferential directions.

We choose the least restrictive unique connection method known to
us: the \emph{Delaunay triangulation}, although one should keep in
mind that this is just one of the many possibilities. There are only
two limitations to the approach presented here: 1) only `neighbouring'
points should be able to communicate; 2) the resultant network, or grid, should
be a simple and connected graph, i.e. two points are connected by at most one line
segment and there is a path from any point to any other point in the graph.
In this way, we create what is called an \emph{unstructured
grid}, to distinguish it from grids with systematic properties such
as cell size or wall direction.

\subsubsection{Voronoi/Delaunay tessellation}

The Delaunay triangulation \citep{Delaunay} and its mathematical dual, the \emph{Voronoi
tessellation} \citep{Voronoi}, is one of the mainstays of \emph{stochastic geometry}.
We will briefly discuss its properties. For more details, we refer
to \citet{Stoyan} and \citet{Okabe}.

A tessellation is an arrangement of polytopes which fit together without
any overlap, completely covering a certain domain. Usually these cells
are convex, which means that every line connecting any two points
within the cell is also within the cell. A very important and exhaustively
studied category of tessellations is the \emph{Voronoi tessellation}.
It is widely applicable in numerous branches of theoretical and applied
science, from astrophysics to zoology \citep[e.g.][]{Weygaert}.

Given a stationary point process \( \Phi  \), of nuclei \( \{x_{i}\} \)
in \( \mathbb{R} ^{d} \), which has a finite intensity \( n  \),
the Voronoi tessellation is defined as 
\begin{equation}
\label{voronoi_total}
V(\Phi )=\{C_{i}\},
\end{equation}
 in which\begin{equation}
\label{Voronoi_cell}
C_{i}=\left\{ y\in \mathbb{R} ^{d}:\left\Vert x_{i}-y\right\Vert \leq \left\Vert x_{j}-y\right\Vert \forall x_{i}\neq x_{j}\right\}.
\end{equation}
 That is to say, the Voronoi cell \( C_{i} \) is the set of all points
closer to \( \bmath{x}_{i} \) than to all other points.

If two Voronoi cells \( C_{i} \) and \( C_{j} \) have a common \( (d-1) \)-facet
(in two dimensions an edge, in three dimensions a wall, etc.), they
are said to be \emph{contiguous} to each other. By joining all the
nuclei whose cells are contiguous, we obtain a set of simplices (a
\emph{simplex} is the generalisation of a tetrahedron in \( d \)-dimensional
space). Thus, we obtain a second form of tessellation based upon
the same point process. This is the \emph{Delaunay triangulation},
and its simplices are called Delaunay triangles, tetrahedra, etc.

\subsection{Transfer along the Delaunay network}

\subsubsection{Continuous or discrete transfer}

Once a grid has been defined, we must specify a method by which the radiation is supposed to travel along the grid lines.
The usual way is, to compute the entire propagation along each path segment, i.e. to integrate the one-dimensional
version of the equation of radiative transfer. This necessarily entails two problems: first, the necessity of designing
a subgrid model (i.e. an approximation of the optical properties within a computational cell); second, the computer-intensive
effort of calculating this integral for each grid line.

Our approach is different: instead of applying continuous transfer, we move the radiation without further processing from node
to node. Remember that we do not use an underlying grid which is then crossed by the photon characteristics; we dispense with the grid
altogether and use a point-to-point propagation of the radiation. By taking this approach, each Delaunay line is equivalent to each other one.
Of course, this means that the point distribution only \emph{represents} the density, and cannot be directly proportional to the density,
except in the homogeneous case; otherwise, the intensity of a point source would not decrease exponentially in an absorbing atmosphere.
We are therefore obliged to find a suitable mapping between the density distribution of the medium and the discrete points representing it.
We use a local criterion which uniquely determines this mapping: we require the optical mean free path to be locally the same for the 
exact exponential solution and the point-to-point transfer.

Let a particular photon line start at coordinate \( 0 \), and end at a distance \( x \). Assume that our point sampling is so fine
that the density is approximately constant between these points; in other words, \( \frac{1}{\rho}\frac{\partial \rho}{\partial x}<\frac{1}{\lambda _{\rmn{D}}}\),
where \(\lambda _{\rmn{D}}\) is the mean length of a local Delaunay line. Then the radiation arrives at \(x\) with an attenuation \(e^{-x/\lambda}\), where
\(\lambda\) is the photon mean free path. Now we sample the segment \((0,x)\) with \(N\) points, at each of which a fraction \(c\)
of the radiation is taken away. Thus, the discrete propagation attenuation becomes
\begin{equation}
(1-c)^N,
\end{equation}
in which \(c\) is a \emph{global} constant (to be considered below) and in which \(N=x/\lambda _{\rmn{D}}\). To first order, this expression
is equal to the exponential attenuation, if
\begin{equation}
x/\lambda=Nc=cx/\lambda _{\rmn{D}},
\end{equation}
so that
\begin{equation}
\label{temp}
\lambda _{\rmn{D}}=c\lambda.
\end{equation}
The question is which recipe to use for distributing the grid points in such a way that the optical mean free path
\(\lambda\) locally is represented correctly via Eq.(\ref{temp}).

\subsubsection{Placing the points}

We base our point distribution on the local properties of the medium.
From basic radiative transfer theory, we know that \( \alpha ^{\rmn{abs}} = \rho \kappa ^{\rmn{abs}} \)
and \( \alpha ^{\rmn{scat}} = \rho \kappa ^{\rmn{scat}} \), where \( \kappa ^{\rmn{abs}} \) and
\( \kappa ^{\rmn{scat}} \) are the \emph{mass absorption coefficient} and the
\emph{mass scattering coefficient}, respectively. Because the mean free path \(\lambda = 1 / \alpha \),
we know that locally 
\begin{equation}
\label{LambdaToRho}
\lambda(\bmath{x})=1/\kappa\rho(\bmath{x}). 
\end{equation}
Given a local grid point density \(\rho _{\rmn{D}}(\bmath{x})\), we know from stochastic geometry that the average
Delaunay line length \(\lambda _{\rmn{D}}(\bmath{x})\) in that region locally will have length
\begin{equation}
\label{AverageDelaunay}
\lambda _{\rmn{D}}(\bmath{x})=\zeta /\rho_{\rmn{D}}(\bmath{x})^{1/d}, 
\end{equation}
in which \(\zeta\) is some constant geometrical factor, which depends on the dimension \(d\) and has been
evaluated in, for example, \citet{Okabe}, as
\begin{eqnarray}
\zeta_{2D} & = & \frac{32}{9\pi} \approx 1.132\label{2DFactor}\\
\zeta_{3D} & = & \frac{1715}{2304}\left( \frac{3}{4\pi} \right)^{1/3} \Gamma\left( \frac{1}{3} \right) \approx 1.237\label{3DFactor}.
\end{eqnarray}
We can conclude from Eqs.(\ref{LambdaToRho}) and (\ref{AverageDelaunay})
that, if we choose our point distribution to sample the \(d\)-th power of the density, i.e.
\begin{equation}
\label{PointDensToDens}
\rho_{\rmn{D}}(\bmath{x})=\frac{\rho^d(\bmath{x})}{\int_{D}\rho^d(\bmath{x})}N, 
\end{equation}
in which \(N\) is the total amount of grid points available and \(D\) the volume of our
computational domain,
the length of a Delaunay
line \( \lambda _{\rmn{D}}(\bmath{x}) \) between two event centres, or grid points, will scale linearly with the \emph{local} mean
free path of the medium \( \lambda(\bmath{x}) \) via a constant \(c\). That is
\begin{equation}
\label{LengthToMFP}
\lambda _{\rmn{D}}(\bmath{x}) = c\lambda(\bmath{x}).
\end{equation}
Thus, because we choose the point distribution to conform to the density profile of the medium
according to Eq.(\ref{PointDensToDens}), the average Delaunay line length and the mean free path
have the same \(\rho^{-1}\) dependence, by which Eq.(\ref{LengthToMFP}) is a \emph{global} relation with
a \emph{global} constant \(c\). We will explicitly derive the important constant \(c\) later on. 
In other words, by adopting the sampling criterion in Eq.(\ref{PointDensToDens}) we have accounted for the difference between integrating
the propagation along a Delaunay line, and using a discrete point-to-point propagation.

Unless we have an enormously high amount of points available, the average local
geometrical mean free path of our graph, \( \lambda _{\rmn{D}} \), will be a lot bigger than
the local physical microscopic scattering mean free path \( \lambda ^{\rmn{scatt}}\). In other words, we expect that
\( c ^{\rmn{scatt}} \gg 1 \). Thus, we will define each of the grid points to be a
scattering centre, where possibly another event can occur, e.g. (partial) absorption.
Of course, \( \lambda ^{\rmn{abs}}\) need not be equal to \( \lambda ^{\rmn{scatt}}\),
but we take care of this by choosing a suitable numerical transfer recipe.

The only problem occurs, when \( \lambda ^{\rmn{scatt}}\) becomes bigger than the
size of our computational domain. In this case, the domain should not contain any
event centres, and should thus be devoid of grid points. We shall tackle this problem
towards the end of this paper. For now, we shall assume that \( \lambda ^{\rmn{scatt}}\)
is smaller than the dimension of our computational domain.

\subsubsection{Propagation}
\begin{figure}
{\begin{center} \includegraphics[width=5cm,clip=]{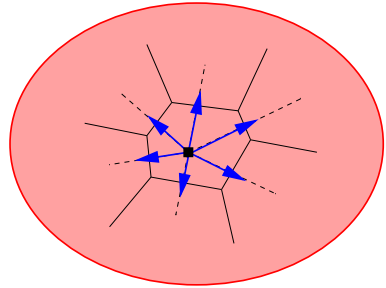} \par \end{center}}
%{\centering \resizebox*{0.5\columnwidth}{!}{\includegraphics{RadProj.eps}} \par}

\caption{\label{AreaProjFig}A blob of radiation (red shaded) is
subdivided in a high number of Voronoi cells, each of which surrounds one grid point.
We magnify one of these Voronoi cells and show how the source radiation is sent along the
dashed Delaunay lines into the neighbouring Voronoi cells.}
\end{figure}

Now let us proceed to radiative transfer on this grid, or graph. From now on we 
will use two-dimensional examples to illustrate the mechanism,
but in every case the generalisation to \( d \)-dimensional space
is either trivial, or else explicitly clarified.
Let us consider the example in which there is a blob of matter which acts as a source
of radiation. According to what was said before, we have to put a number of grid points
within the blob, according to the \(d\)-th power of its density distribution. We know that each
point is surrounded by a Voronoi cell, and we assume that the Voronoi cells are small
enough (i.e. the number of points \(N\) is high enough) to accurately fill the blob, according to some
criterion \( \frac{1}{\rho}\frac{\partial \rho}{\partial x}<\frac{1}{\lambda _{\rmn{D}}}\).
Now we use each of these points as a source, which means that we send an equal
amount of source photons out of this point along each of the Delaunay lines which 
emerge from it. An illustration of this example can be seen in Fig.\ref{AreaProjFig}, in which we 
exaggerated the size of one Voronoi cell. We show a Voronoi cell with its neighbours
and the dashed lines indicate the Delaunay lines.

\begin{figure}
{\centering \includegraphics[width=8.2cm,clip=]{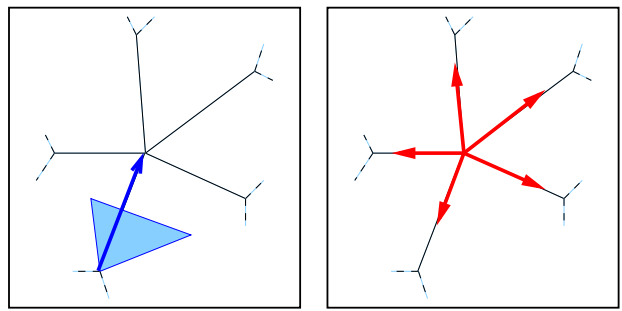} \par}
%{\centering \resizebox*{0.5\columnwidth}{!}{\includegraphics{IntersectionNew.eps}} \par}

\caption{\label{GridPointSplit}\emph{Left}: The radiation in the shaded
area is projected onto the Delaunay edge and propagates along it.
\emph{Right}: When it comes upon an intersection, it is split up according to a
certain recipe which depends on the events taking place at this
event centre.}
\end{figure}

Thus, the whole domain is subdivided unambiguously by the Voronoi cells and the radiation
is projected onto the Delaunay network. The only thing left to be specified
is a way to let the source radiation propagate solely along the resultant Delaunay line network,
from one event centre to the next. Zooming in on one grid point (see Fig.\ref{GridPointSplit}, left), which
is connected to a number of others, we see that the source radiation within
the shaded area, which is a certain part of the original Voronoi cell,
is projected onto the Delaunay edge. It propagates
along that edge until it reaches the next point, where a number of
Delaunay lines meet. Because this grid point is a scattering centre,
the radiation will be sent away out of the cell along all the Delaunay
line (see Fig.\ref{GridPointSplit}, right), and propagates onwards
along the graph until it reaches the next event centre where we can
use the same general recipe, because of the Markovian properties of
this random walk. Moreover, we are at liberty to put in 
all kinds of radiation-matter interactions (e.g. absorption, re-emission,
multi-frequency redistribution, ionisation, recombination) between the arrival 
and scattering of the radiation packet, but we will elaborate on this later. 

Thus, we have split the radiative transfer equation into two parts: 1) we let
the radiation interact with the medium at each event centre, after which 2)
we advect the radiation along the Delaunay lines towards the next event centre.
By making this choice, we have reduced the radiative transfer to its
microscopic origin: a random walk of photons between scattering centres.
As said, the physical mean free path of photons is not the same
as the mean length of our Delaunay lines, so this Markov process does
not coincide with the microscopic physical case. Even so, we believe that
it retains the essentials (cf. Eq.(\ref{LengthToMFP})), while removing the grid-dependent systematic
effects mentioned above in the previous section.

\subsection{A physical grid}

\begin{figure}
{\includegraphics[width=8.1cm,clip=]{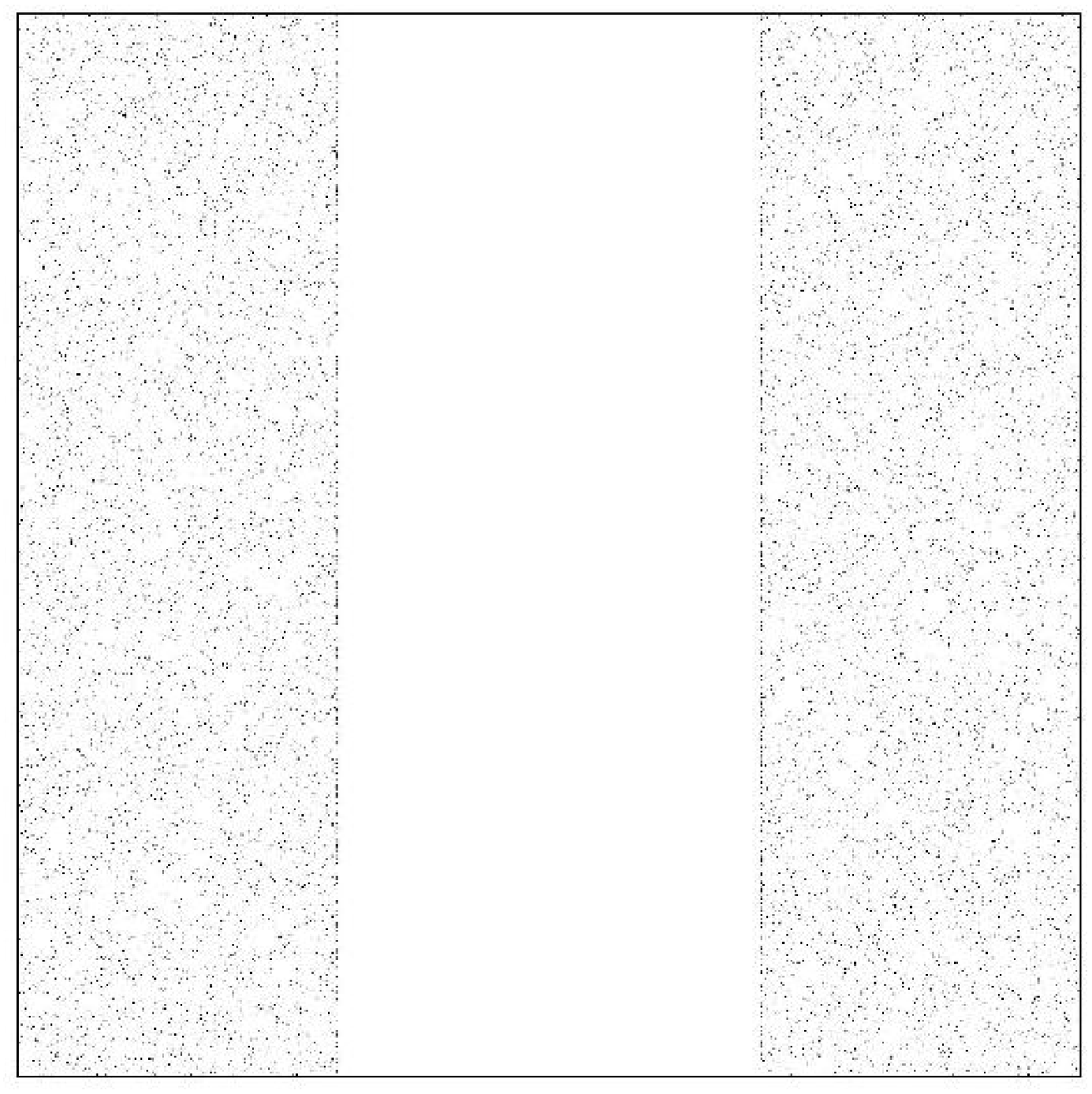}
\includegraphics[width=8.1cm,clip=]{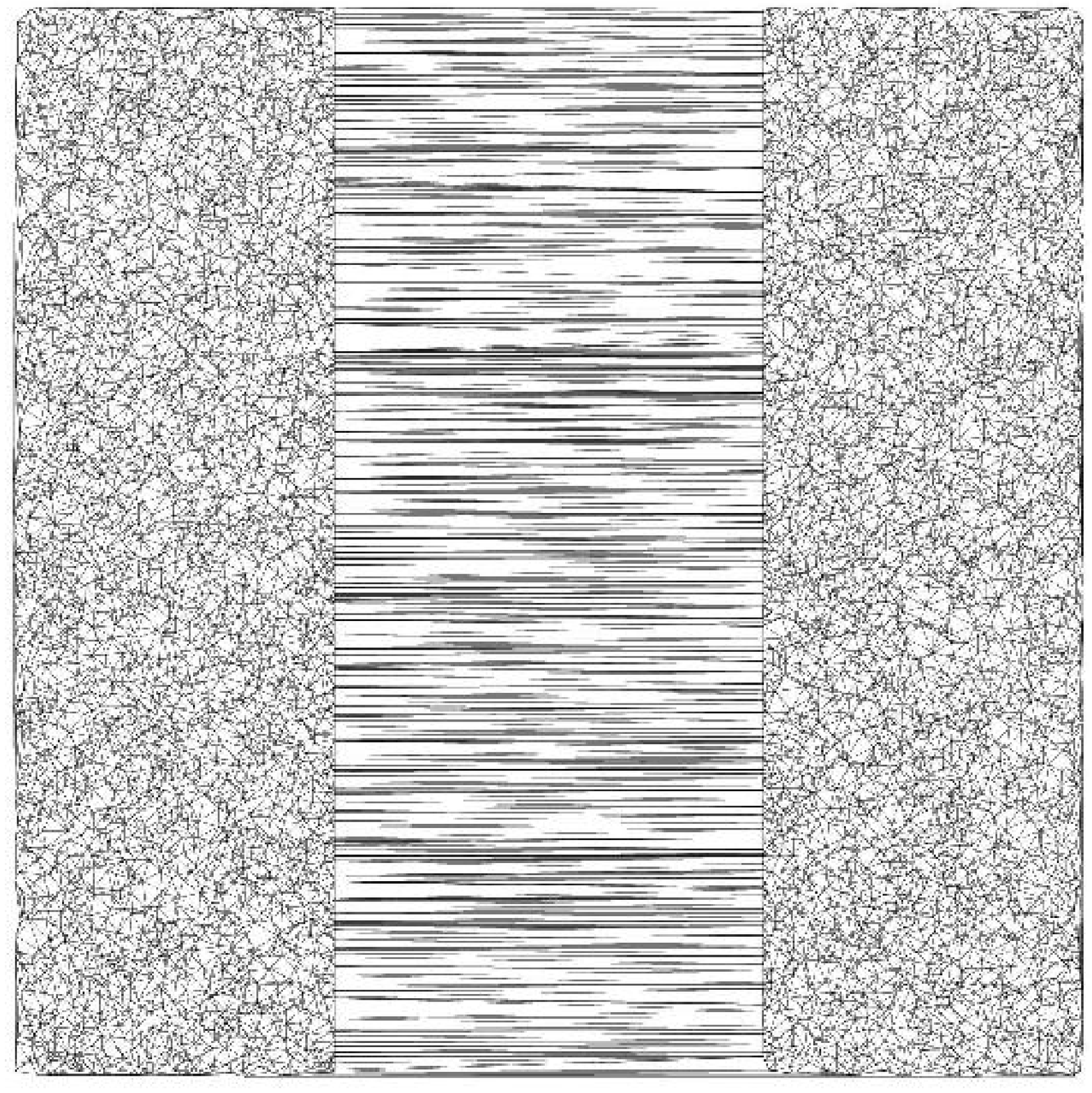}}

\caption{\label{CloudsDel}\emph{Top}: Point distribution for a medium distribution
as in Fig.\ref{CloudsZoom}. \emph{Bottom}: Resultant Delaunay network of characteristics
along which the radiation can propagate.}
\end{figure}

To illustrate the optimal physical properties of the geometry of the grid, we mention
again that by choosing the point distribution properly the average length of a Delaunay line, and thus the average
width of a Voronoi cell, scales linearly with the local mean free path
of the photon (cf. Eq.(\ref{LengthToMFP})), by which the scale length is
not some superimposed unphysical measure, but directly related to the physical
properties of the medium. Moreover, because we use a stochastic point process
to place the grid points, the angle between two lines meeting at one of these
points will also have a stochastic nature, which removes the unphysical fixed
preferential directions superimposed by the numerical methods mentioned in
Section 2.
In order to show an example of the type of grid we obtain by using our method, we return
to the example used in Fig.\ref{CloudsZoom}. Using our recipe for placing grid points
according to the medium density profile, we obtain the result depicted in Fig.\ref{CloudsDel}, top.
In the bottom part of Fig.\ref{CloudsDel}, one can see the result of making
a Delaunay tessellation of this point distribution. One can readily see that
the length of the Delaunay lines, that is the characteristics along which the
radiation can propagate, correlates with the mean free path. The Delaunay lines
emerging from the grid points at the boundary of cloud I are very long and causally connect
cloud I with cloud II. The Delaunay lines are precisely the emergent perpendicular
characteristics of the plane wave moving through the optically thin region.
Because nothing happens to the radiation packet as it moves in a straight line
along these characteristics to the first event centres in cloud II, the
numerical diffusion is minimal compared to what we would get using a short characteristic
method.

Another advantageous property of a Delaunay network is that it increases
the angular resolution in cases where it is needed. If we would have a
homogeneous medium distribution full of scattering centres that are distributed
according to a Poisson point process, the resultant Delaunay graph has some
properties that can be evaluated analytically \citep[for a summary, see][]{Okabe}.
One of these is the average number of \( d-1 \) facets of a \( d \) dimensional
Poisson-Voronoi cell, that is the number of sides in two or the number
of walls in three dimensions. These are \( 6 \) and \( (48\pi^2/35)+2 \approx 15.54 \),
respectively. The normal of each \( d-1 \) facet is a Delaunay line, thus
the number of \( d-1 \) facets of each Voronoi cell corresponds to the
number of Delaunay lines joining at an event centre or, equivalently, the
amount of (solid) angles into which the radiation can be scattered (see Fig.\ref{GridPointSplit}).
Because the possible directions for the homogeneous medium is on average \( 6 \) 
(in 2D), the angular resolution is not very good. But this poses
no problem, because, as we argued in Section 2, in this case of a high scattering
opacity the angular dependence tends to get wiped out. 
The angular resolution \emph{is} of importance, when radiation is sent out
into an optically thin region. Therefore, the angles between the Delaunay lines 
connecting a grid point near the boundary of cloud I to one in cloud II have to be small.
We will more rigorously show in a later section of this paper that a Delaunay
tessellation based on a stochastic inhomogeneous point process has the property
that the angle between long Delaunay lines will be much smaller than the angle
between short Delaunay lines.

Another elegant property of our grid is that it is reciprocal, or 
time-reversible. In a short-characteristics scheme, we need to devise
a clever downwind sweeping and interpolation scheme to know the influence
of an upwind source A on a downwind point B (cf. Fig.\ref{CloudsZoom}). If we want to turn things
around, and try to calculate the influence of point B on point A, the
overall cascade of computations will, in general, differ. In a long characteristic
method, we would create a high number of isotropically distributed rays at point A, one
of which hopefully passes nearby point B, but one would need to create a whole new
set of rays around B to calculate its effect on cloud I.
Our grid has an inherent symmetry, in that we know that the trajectory of a photon
packet sent out from point A across the network to point B will be the same for a 
photon sent out from point B to point A. This is the best example of what we
already mentioned earlier, that our grid is not designed for one specific
geometrical symmetry, as so many of the extant numerical methods, but that
it is symmetrically optimised for each grid point.

\subsection{Event centres}

As we have pointed out earlier, our new method splits the radiative transfer
equation into two parts: one advection part, and one interaction part. It's
at the event centres, or grid points, where the radiation-matter interaction
takes place. In the previous, we defined each event centre to be at least
a scattering centre, where the radiation is redistributed according to the
recipe in Fig.\ref{GridPointSplit}, right. But our method gives us the liberty
to incorporate a wide variety of radiation-matter interactions in between
the arrival and scattering of the radiation, such as absorption, ionisation, 
re-emission, recombination, etc.

\subsubsection{Absorption}
Suppose one wants to simulate the propagation of photons through an absorbing
medium, or, more specifically, one wants to determine the radiation intensity profile
as a result of a certain distribution of sources in a medium with a mass absorption
coefficient \( \kappa ^{\rmn{abs}} \). If this medium is inhomogeneous, the
density can change dramatically from one point to the next, so that the value of the absorption
coefficient (via \( \alpha ^{\rmn{abs}} = \rho \kappa ^{\rmn{abs}} \)) can have widely
varying values at different places. Because we have designed our method in such a way
that our point distribution follows the medium distribution according to Eq.(\ref{PointDensToDens}), we can assign a \emph{global constant}
to each grid point.

\begin{figure}
\begin{center}
\includegraphics[width=4.1cm,clip=]{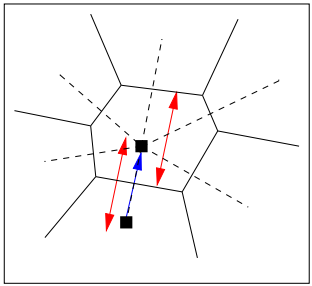}

\caption{\label{TauFig}The incoming radiation (blue arrow) sees an optical depth (e.g. for
absorption) that is equal to the local absorption coefficient times the average width of the
Voronoi cell. To first order, the length of the incoming Delaunay line is equal
to the width of the Voronoi cell (see red arrows).}
\end{center}
\end{figure}

What is the value of this constant? Let us examine Fig.\ref{TauFig}, in which a typical Voronoi
cell which surrounds an event centre is depicted. Radiation comes in from a neighbouring
Voronoi cell along a Delaunay line (blue arrow), and, before it is scattered, some part
of it is absorbed according to the local optical depth \(\Delta\tau^{\rmn{abs}}\), which we choose to be equal
to the local opacity \(\alpha^{\rmn{abs}}\) at \emph{this} grid point times the length of the 
incoming Delaunay line, which is, to first order, equal to the average width of the Voronoi
cell (see Fig.\ref{TauFig}).
Thus, \(\Delta\tau^{\rmn{abs}} = \alpha^{\rmn{abs}}\lambda_{\rmn{D}}\) and, upon using Eq.\ref{LengthToMFP}, 
we have \(\Delta\tau^{\rmn{abs}} = c^{\rmn{abs}}\), which is the 
number of mean free paths contained in the Delaunay length \(\lambda_{\rmn{D}}\). So, if we assign a constant
\(c^{\rmn{abs}}\),
representing the local optical depth, to each grid point, the fraction \(1-\rmn{e}^{-c^{\rmn{abs}}}\) 
of the incoming radiation is attenuated at this event centre due to absorption.  
Assuming that the length of a Delaunay line is much smaller than the absorption mean free path 
(otherwise the radiation cannot travel far), or equivalently \(c^{\rmn{abs}}\ll1\), we can expand
the exponential to first order: \(1-\rmn{e}^{-c^{\rmn{abs}}}\approx c^{\rmn{abs}}\).

By this, we can incorporate absorption in our method by assign a constant \(c^{\rmn{abs}}\) to each grid point,
and by defining an absorption recipe:
\begin{eqnarray}
I^{\rmn{abs}} & = & I^{\rmn{in}}c^{\rmn{abs}} \label{AbsRecipeAbs}\\
I^{\rmn{out}} & = & I^{\rmn{in}}(1-c^{\rmn{abs}})\label{AbsRecipeOut},
\end{eqnarray}
in which \(I^{\rmn{in}}\) is the amount of incoming radiation (blue arrow in Fig.\ref{TauFig}),
\(I^{\rmn{abs}}\) is the amount of radiation locally absorbed, and \(I^{\rmn{out}}\) is the amount
of radiation which will leave the event centre according to the scattering recipe in Fig.\ref{GridPointSplit}.
Thus, we have circumvented the inherent difficulties of the extant numerical methods (interpolating and
integrating optical depths, evaluation of exponentials, etc.) and reduced the whole absorption process
to the computationally efficient calculation of fractions.

We conclude by noting that \(I^{\rmn{in}}=I^{\rmn{out}}+I^{\rmn{abs}}\) always, by which our method
is explicitly photon-conserving, in contrast to other methods which lose photons due to interpolations and other systematic errors.

\subsubsection{Other processes}
Of course, the recipe in Eqs.(\ref{AbsRecipeAbs}) and (\ref{AbsRecipeOut}) can be used
for various other radiation-matter interactions. If we have, for example, a mixture 
of two or more different gases, which have the same overall density profile, but a different
mass absorption coefficient, we would have several different \( c^{\rmn{abs}} \)'s assigned to 
each grid point, one for each type of gas.

Another useful physical process we can easily incorporate into our method is ionisation,
for which we can use the local optical depth \(c^{\rmn{ion}}\) as a measure of the amount of neutral
atoms at each event centre. It is important that our method \emph{is} photon-conserving, because
this ensures the right dimensions of a resultant Str\"omgren sphere, or, more accurately in an inhomogeneous
medium, Str\"omgren \emph{region}.

Furthermore, radiation that was absorbed can be re-emitted by adding it to the outgoing radiation locally,
maybe even in a different frequency domain. Frequency dependence can easily be introduced by, for example,
introducing an interaction matrix \(A_{ij}\) which determines the redistribution of radiation \(I_i\) from one frequency domain \(i\)
into radiation \(I_j\) in another domain \(j\) according to the local physical properties of the medium.
These physical properties may even include complicated chemical equilibrium networks, or feedback effects from
some independent hydrodynamical solver. Even Doppler effects can be taken into account.

\subsection{Interaction coefficients}
Now that we have shown that we can incorporate any local radiation-matter effect into our method
by attaching several constants, one for each kind of interaction, to each grid point, we still have
to specify how to determine the value of the constants in the set radiation-matter interaction coefficients \(\{c^i\}\), 
which we assign to each grid point.

To do this, we proceed as follows: given a medium distribution profile in the form of a scalar function (or, as mentioned
before, array) \(\rho(\bmath{x})\), in which \(\bmath{x} \in D=[0.0:1.0]^d \) which is the size of our 
\(d\) dimensional computational domain, we have distributed our \(N\)
available grid points in such a way that it accurately samples the function \(\rho^d(\bmath{x})\). Of course,
we can choose our point distribution to follow a different function of the density, \(f(\rho(\bmath{x}))\), but
Eq.(\ref{LengthToMFP}) is only valid \emph{globally} with a constant \(c\), when \(f(\rho(\bmath{x}))=\rho^d(\bmath{x})\). 
Because Eq.(\ref{LengthToMFP}) is valid in the \emph{whole} medium, it is
also valid at some location \(\bmath{x}_0\) where the medium density is equal to its average density, that is at a location
where
\begin{equation}
\label{AverageDensity}
\rho(\bmath{x}_0)=\left\langle\rho(\bmath{x})\right\rangle=\int_D \rho(\bmath{x})\rmn{d}x=M_{\rmn{tot}},
\end{equation}
in which \(M_{\rmn{tot}}\) is the total mass of the medium inside the computational domain.
Given the mass `interaction' coefficient \(\kappa^i\) for a certain radiation matter interaction, the
mean free path at \(\bmath{x}_0\) is
\begin{equation}
\label{AverageMFP}
\lambda(\bmath{x}_0)=1/\kappa^i M_{\rmn{tot}}.
\end{equation}
Because the value of the medium density at \(\bmath{x}_0\) is the expectation value, the value of
the grid point density at \(\bmath{x}_0\) is also the expectation value. Thus, because the volume
of our computational domain is unity,
\begin{equation}
\label{AveragePointDens}
n(\bmath{x}_0)=N,
\end{equation}
where \(N\) is the number of available grid points, that is our resolution. As said, we know
from stochastic geometry that the local Delaunay line length is determined by Eq.(\ref{AverageDelaunay}).
Combining all these elements, we get
\begin{equation}
\label{cidetermined}
c^i=\frac{\zeta M_{\rmn{tot}}}{N^{1/d}}\kappa^i.
\end{equation}

Thus, given our resolution \(N\) and the total mass \(M_{\rmn{tot}}\) of the medium inside our computational
domain, Eq.(\ref{cidetermined}) determines the global constant that we have to attach to each grid point
for a certain radiation-matter interaction characterised by a mass `interaction' coefficient \(\kappa^i\).

It is, of course, possible that \({\kappa^i}\) is not a constant, but is a function of, for example, the local temperature. 
In that case, Eq.(\ref{LambdaToRho}) is no longer generally valid and, in order to make sure that Eq.(\ref{LengthToMFP}) still holds, we must scale
the point density to the more general opacity function:
\begin{equation}
\rho_{\rmn{D}}(\bmath{x})=\frac{\alpha^d(\bmath{x})}{\int_{D}\alpha^d(\bmath{x})}N.
\end{equation}
However, we use the scaling relation Eq.(\ref{PointDensToDens}) whenever we can, because in most cases densities
are the relevant quantities obtained from hydro-solvers.

We make a final note that, if we choose a point distribution dependent on the density in a form different from Eq.(\ref{PointDensToDens}),
Eq.(\ref{LengthToMFP}) is still valid locally, but now with a spatially varying \(c^i(\bmath{x})\) in the form
\begin{equation}
\label{cispatdependent}
c^i(\bmath{x})=\frac{\zeta \rho(\bmath{x})}{\rho_{\rmn{D}}(\bmath{x})^{1/d}}\kappa^i.
\end{equation}
Of course, it is much easier to have just one global constant \(c^i\) for each grid point, which is why we prefer
to use a point distribution in the form of Eq.(\ref{PointDensToDens}), but this is not mandatory at all. In fact,
if we were to extend our transport method to higher dimension (\(d>3\)), which is possible in applications beyond
radiative transfer (e.g. data streams in \(d\)-dimensional space), the equivalent of Eq.(\ref{PointDensToDens}) would
assign too many points to just a few regions in \(d\)-space; in which case a varying value of \(c^i\) would be preferable.

\subsection{Resolution issues}
Because we use a recipe Eq.(\ref{PointDensToDens}), our point distribution conforms to the features of the
medium density profile. Therefore, it makes no distinction between optically thin and optically thick regions,
and the same recipe (variants of Eqs.(\ref{AbsRecipeAbs}) and (\ref{AbsRecipeOut}) in combination with the scattering recipe in Fig.\ref{GridPointSplit}) 
for the whole medium. But, because the contrast in medium density can, in realistic cases, be extremely high,
and because this contrast is exaggerated by an exponent \(d\) by using a point distribution recipe Eq.(\ref{PointDensToDens}),
we need a way to reduce the overabundant resolution in high density areas, so that we can use that part of
our finite amount of available grid points \(N\) in places that are undersampled.

A way of doing this is by cutting off the grid point distribution function \(\rho_{\rmn{D}}(\bmath{x})\) at some user-specified
value, e.g. at a factor \(f>1\) above the average point density, that is at \(n_{\rmn{max}}=f\left\langle \rho_{\rmn{D}}(\bmath{x}) \right\rangle\), and
locally replace the \(k\) overabundant grid points by, for example, just one. Of course, this means that the
constants \(c'^i\) that are to be assigned to this one point are different from the global constants \(c^i\) as
evaluated in Eq.(\ref{cidetermined}). Because this point now represents \(k\) separate ones, the recipe in 
Eqs.(\ref{AbsRecipeAbs}) and (\ref{AbsRecipeOut}) has to be used \(k\) times at this one point. So, given a
global absorption constant \(c^{\rmn{abs}}\), a fraction \((1-c^{\rmn{abs}})^k\) of the incoming radiation remains to
be emitted. Thus, we get the same overall behaviour, if we define a new local constant \(c'^i\)
\begin{equation}
\label{ciheavypoint}
c'^i=1-(1-c^i)^k.
\end{equation}
If \(c^i\ll1\), \((1-c^i)^k\approx 1-kc^i\), by which \(c'^i\approx kc^i\), but this introduces a error, albeit small,
so we stick to the recipe in Eq.(\ref{ciheavypoint}).

\subsection{The whole algorithm}
Now that we have introduced all the basic ingredients of our new numerical method, we can describe the full
recipe. Our whole simulation can be split up into \(S\) individual equivalent steps, each of which consists of
the following combination of ingredients:
\begin{enumerate}
\renewcommand{\theenumi}{(\arabic{enumi})}
\item Collect all the (multi-frequency) radiation that is emitted by this event centre (e.g. it is a source, or radiation
is re-emitted), and all the radiation that arrives via the incoming Delaunay lines;
\item Use recipes in the form of Eqs.(\ref{AbsRecipeAbs}) and (\ref{AbsRecipeOut}) to let this radiation interact with
the medium;
\item Send out the resultant (spectrum of) radiation onto the Delaunay network according to the scheme depicted in
Fig.\ref{GridPointSplit} until it hits the next event centre. If the emission and/or interaction is anisotropic, 
one can choose to distribute it onto the lines accordingly.
\end{enumerate}
Starting with a list of \(N\) randomly ordered grid points, we define one step as applying this list of actions once
for every grid point.

Focusing on just one packet of source radiation emitted at a grid point in the middle of the computational domain, we
can analytically estimate how far it will spread along the Delaunay network. Assuming that the \(N\) points are distributed
homogeneously and have a set of interactions coefficients \(\{c^i\}\) attached, we can derive from basic random walk theory that 
after \(S\) steps the second order expectation value of the
displacement \(\bmath{R}\) of the radiation packet has the form
\begin{equation}
\label{Displacement}
\left\langle \bmath{R}^{2}\right\rangle = \lambda_{\rmn{D}} ^2S,
\end{equation}
in which \(\lambda_{\rmn{D}}\) is determined by Eq.(\ref{AverageDelaunay}). Therefore, the root mean square of the \emph{net} displacement is
\begin{equation}
\label{NetDisplacement}
\bmath{R}_{\rmn{rms}}=\lambda_{\rmn{D}}\sqrt{S},
\end{equation}
Here, we have assumed that the scattering at each grid point is isotropic, but the same derivation will also hold for any
scattering with front-back symmetry, as in Thomson or Rayleigh scattering.

After \(S\) steps, the total intensity of the packet will have diminished by a factor \(\rmn{e}^{- S\sum _{i} c^i}\). If the number of steps
is high enough and if the mean free path of the photon packet is large enough, the photon will reach the boundary, where it
is absorbed, reflected or leaves the computational domain, depending on what kind of boundary we choose. Often,
the radiation will be fully absorbed long before it reaches the boundary.

In \(d\) dimensional
space, the number of steps needed for the radiation packet to cover the domain, that is, to reach the boundary, is of order 
\( O\left( N^{1/d} \right) \), and at each step the number of operations in the form of Eqs.(\ref{AbsRecipeAbs}) and (\ref{AbsRecipeOut}) scale
with the number of grid points, i.e. are of order \(O(N)\), by which we expect that the simulation will have converged after an operation count
of order \(O\left( N^{1+1/d} \right) \). Thus, the operation count of our method is independent of the number of sources, which
makes it possible to accurately simulate the radiation field of large extended sources just as rapidly as for just one or two point sources.
Another thing we can conclude is that our method has a lower operation count, if the dimension \(d\) is higher, but this is of no
importance, because we need more points \(N\) to accurately sample a higher-dimensional domain. Finally, we should mention that, if we 
include frequency-dependence, characterised by \(N_{\nu}\), the operation count will be of order \(O\left( N_{\nu} N^{1+1/d} \right) \).

\section{Implementation}

Having shown how our method works and what its mathematical properties are, we
are in the position to show what results we get from implementing the method.
We first test the method on a grid based on a Poisson point distribution, 
because we know that, because of Eq.(\ref{LengthToMFP}), the homogeneous distribution
is just a (special) type of inhomogeneous distribution, and the two cases are
equivalent, algorithm-wise. In the next section, we will use the example
of a special correlated point process which results in an inhomogeneous point
distribution.

We use the algorithm
described in \citet{Barber} to construct the Delaunay
triangulation. It has been proven to perform the tessellation in \( O(N\log N) \)
expected time for \( d\leq 3 \), and in \( O\left( N^{\left\lfloor d/2\right\rfloor }\right)  \)
expected time for fixed \( d\geq 4 \). The source code of an excellent and
much-used implementation is freely available at \texttt{http://www.qhull.org}.

\subsection{Point source}

The first test case we used was a static point source radiating isotropically
in the centre of the \( [0.0:1.0]^2 \) domain.

\subsubsection{Expanding wave-front}

\begin{figure}
{\begin{center} \resizebox*{0.95\columnwidth}{!}{\rotatebox{90}{\includegraphics{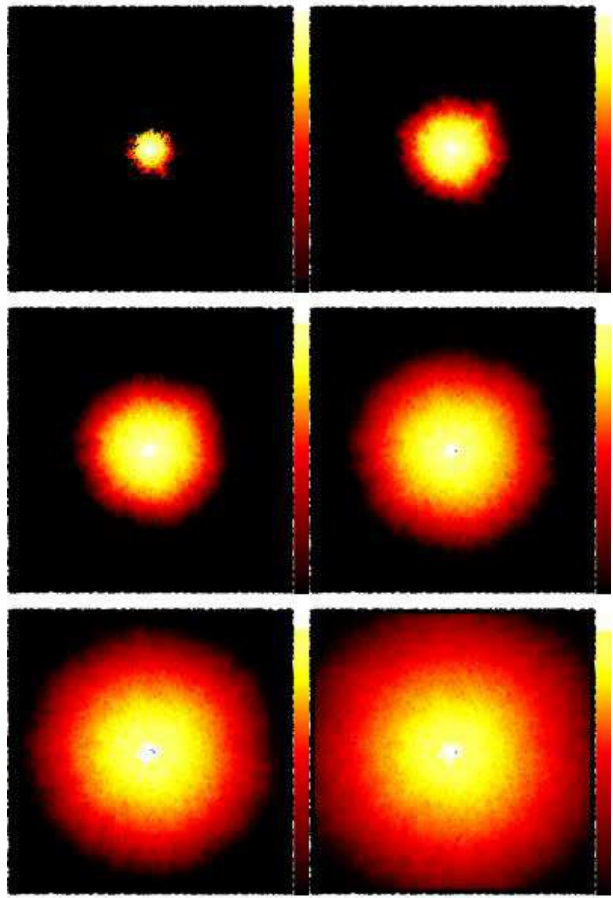}}} \par\end{center}}

\caption{\label{diffusion_fig}Illustration of the wave-front expansion,
in which a point source is used on a homogeneous grid made up
of 50000 points. Shown is the logarithm of the intensity distribution after (left
to right; top to bottom) 3, 9, 15, 25, 35 and 50 steps.}
\end{figure}

As we have described in the previous section, at each step the radiation is 
split up and redistributed once at every grid point, by which the radiation
can propagate only as far as one edge length along the grid. 
In Fig.\ref{diffusion_fig} we plot the logarithmic
intensity of the radiation within each Voronoi cell for an increasing
number of steps (note that we use an absorbing boundary). We
determine the radiation within one Voronoi cell by averaging the radiation
arriving, via the (on average, six) Delaunay lines, at the cell's
nucleus. The solution is scale-free, so we do not need to specify
the power of the source. But we do note, that the scales next to
each of the six graphs have the same maximum and minimum values.

Comparing this process of wave-front propagation with the mathematical random walk
analysis we gave in the previous section, we see that our process indeed converges
to a stable solution in a finite number of steps of the order \( O\left( N^{1/d} \right) \),
given that \(N=50000\). Indeed, we find that the difference
between the result after \( 50 \) (see Fig.\ref{diffusion_fig},
bottom right) and \( 51 \) steps is negligible. Thus,
this implementation shows that performing
several steps will make the simulation tend to the static one
within polynomial time.

\subsubsection{Absorption profile}

We already mentioned in Section~3 that we can mimic, for example, the absorption
profile by withholding a certain amount of radiation at each intersection, according
to the local optical depth, or interaction coefficient, \(c^{\rmn{abs}}\).
Given this test case example of a point source in the centre of our domain,
we can explicitly examine how the intensity profile changes as we vary the value
of \( c^{\rmn{abs}} \).

\begin{figure}
{\begin{center} \includegraphics[width=8.4cm,clip=]{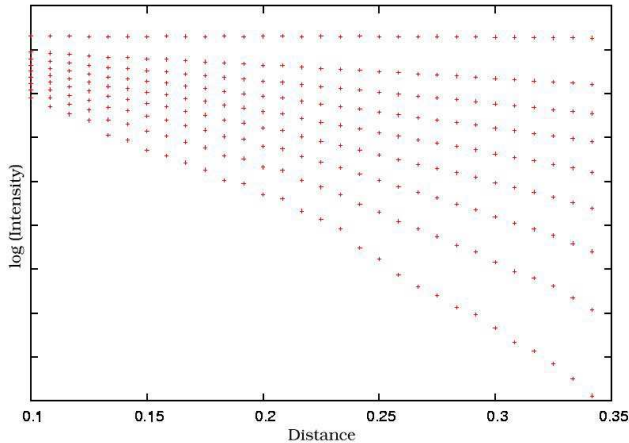} \par\end{center}}
%{\centering \resizebox*{1\columnwidth}{!}{\includegraphics{Abs_CoeffNew.ps}} \par}

\caption{\label{absorption_fig}Plot of the logarithm of the intensity in
each shell versus the distance of that shell to the point source for
(top to bottom) \( c^{\rmn{abs}}=0.0,0.1,..,0.8 \).}
\end{figure}

We subdivide the domain using thirty shells of equal width, concentric
about the radiation source, and compute the amount of radiation within
those shells for various \( c^{\rmn{abs}} \). The results for \( c^{\rmn{abs}}=0.0,0.1,..,0.8 \)
can be seen in Fig.\ref{absorption_fig}, in which the logarithm of
the intensity is plotted versus the distance of the shell to the centre
of the domain (of size \( [0.0:1.0]^2 \)). 

In the absence of absorption, the integrated radiative flux through
each circle concentric about the source is constant. In order to check this, 
we examined the difference of the flux through adjacent circles at each step of the simulation. 
After a few steps, the flux became constant. In the presence of absorption, the solution
to the transfer equation is (for \( d\geq 1 \))
\begin{equation}
\label{intensity_drop}
I(r)=I_{0}\rmn{e}^{-\alpha r},
\end{equation}
 in which \( r \) is the distance to the point source and \( \alpha  \)
is an absorption coefficient. Splitting up the domain in concentric
circular shells as above, we may verify that the total amount of energy
in each ring obeys Eq.(\ref{intensity_drop}). This is indeed the
case.

More specifically, at a distance \(r\), the amount of grid-points encountered
is on average \(r/l\), by which the source intensity \(I_0\) has reduced by
a factor \((1-c^{\rmn{abs}})^{r/l}\). From elementary calculus, we know that, if 
the number of steps, or grid points, \(r/l\) is high, we obtain
\begin{equation}
\label{limit_intensity_drop}
\lim_{r/l\rightarrow\infty}(1-c^{\rmn{abs}})^{r/l}=\rmn{e}^{-rc_i^{\rmn{abs}}/l}=\rmn{e}^{-r/\lambda^{\rmn{abs}}},
\end{equation}
in which we used Eq.(\ref{LengthToMFP}) for the last equality. Thus, we see that
Eq.(\ref{intensity_drop}) and (\ref{limit_intensity_drop}) match, when the number
of grid points \(r/l\) or, more general, \(N\) is large enough.

\begin{figure}\begin{center}
{\includegraphics[width=7.5cm,clip=]{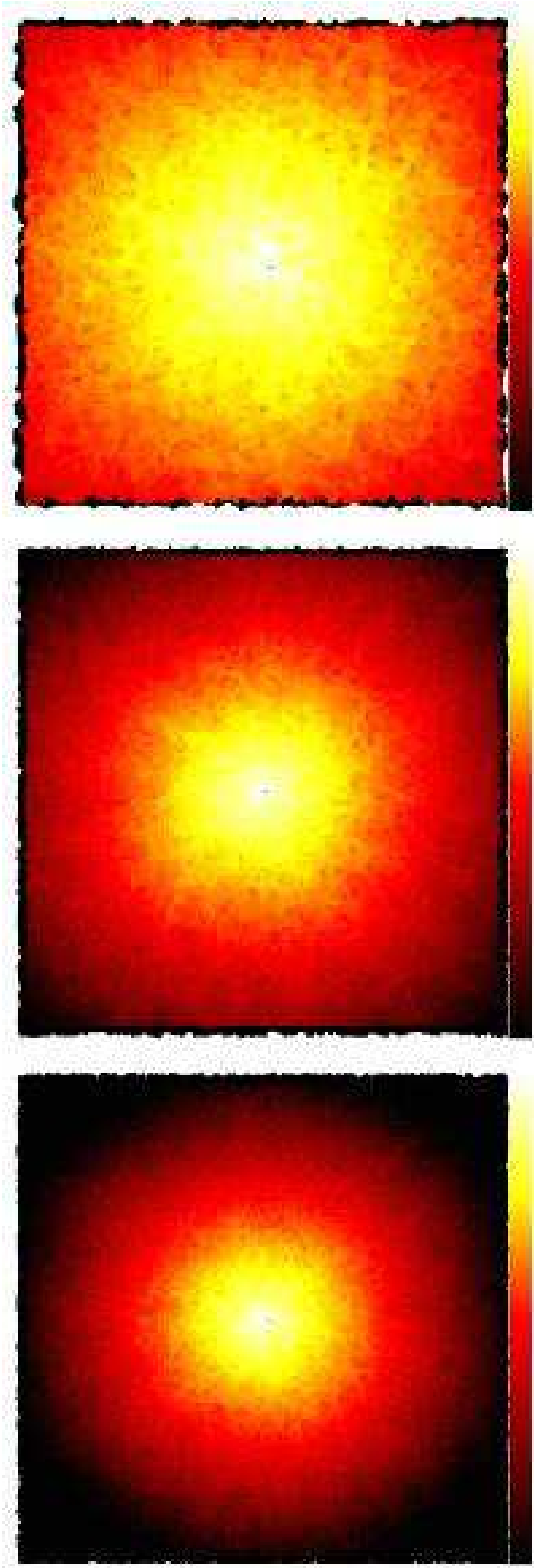}}
%{\resizebox{\hsize}{!}{\includegraphics{StagesRot.ps}}}

\caption{\label{resolution_fig}Converged results for a grid of (\emph{top}
to \emph{bottom}) 10000, 20000, and 50000 points.}
\end{center}
\end{figure}

As an extra check, we computed the slopes for various
values of \( N \), and found that the slope indeed steepened with
a factor \( \left( N_{new}/N_{old}\right) ^{1/d} \), as is to be expected
from Eq.(\ref{cidetermined}). Using more points
simply leads to more absorption, as can be seen in Fig.\ref{resolution_fig},
in which the converged results for the same point source are plotted
for \( N=10000 \), \( N=20000 \) and \( N=50000 \). We used the
same \( c^{\rmn{abs}} \) for each one. In the \( N=10000 \) plot, one can still
discern the individual Voronoi cells.

\subsection{Line source}

We performed all of the above tests in the case of a line source.
In our case, we put the static source at the right boundary of the
grid. Similar considerations as those in Subsection~4.1.2 lead us
to the conclusion that the intensity profile should conform to the
result Eq.(\ref{intensity_drop}). Because of the change of symmetry
from a rotational to a mirror one, the shells should now be straight
parallel strips of equal width. The remainder of the results is identical
to those that are shown in Fig.\ref{absorption_fig}.

\section{Clumpy medium distribution}

Computation of the transfer of radiation as a Markov process on a
Delaunay grid converges to the analytical solution, and an implementation
of our method works well with several test cases.

To this, we wish to add the following considerations. First, we already
stressed that one of the key issues of our method is that we make
use of a stochastic point process to position the available grid points.
If we construct a grid from these points, the angles between the grid
lines also have a stochastic nature. In this respect, our method shares
some characteristics with Monte Carlo methods. Second, if we were
to increase the number of grid points to infinity, we would have enough
grid points to actually sample all medium particles, and our Markov
process would reduce to the actual physical one.

Another important issue is that we did not use the fact that we could
enhance the accuracy obtained, when using a fixed number of points
\( N \), by one-dimensionally solving the transfer equations along
the grid lines. As we argued in Sect. 3.2.1, we have not used this procedure for two reasons. First,
because our method is computationally very efficient, we can just
use a large \( N \) in order to suppress the statistical noise, making
the results more accurate without introducing computational complexities.
Second, in the implementation presented here, the length of a grid line is of no importance for our numerical
scheme. This enables us to draw the conclusion (Sect.4) that, if the
results are correct in the case of a homogeneous point distribution,
they will automatically be correct also in the case of a correlated
distribution. Furthermore, using the scaling in Eq.(\ref{PointDensToDens}) instead of solving
the transfer equation along a Delaunay line produces an enormous speedup of the computation.

Because the length of a grid line is not used numerically, we can
adjust it without changing the result, as long as we make sure that
the point distribution represents the medium distribution. So, if we rescale
all grid lines to the same constant length, and if we make sure that
the medium rescales accordingly, such that a grid point remains attached
to the same patch of medium, we do not only obtain a homogeneous point
distribution, but also a homogeneous medium distribution, for which
we know the method works, according to the tests in Section~4. The
essential issue is that the numerical calculations to be performed
on the grid will still be exactly the same as they would have been
on the original grid, because of the length scale invariance. The connectivity
of a "homogenised" inhomogeneous grid will, of course, be different, but
that does not influence the convergence of our method.

\subsection{Correlated point processes}

So far, we have only used homogeneous (Poisson) point processes, which
represent homogeneous medium distributions. It is straightforward
to solve the radiative transfer equations analytically in this regime
of constant absorption coefficients. Using our method in this case
will solve the transfer equations fast and accurately, but its performance
will not differ significantly from other methods which are able to
solve the transfer equation under similar conditions. The main advantages
of our method are best pointed out when using a correlated point process,
which mimics a clumpy medium distribution. Such clumpy distributions
produce steep gradients in, for example, the absorption coefficient, when the radiation
propagates outward from regions of tenuous medium to dense regions
with a very high opacity. It is in this regime, in which it is impossible
to find an exact analytical solution to the transfer equations, that
it will be difficult and computationally costly to use the regular
numerical schemes as discussed in Section 2.2 for solving the equations.
Each one needs a superimposed grid, and because the size of a grid
cell should be small enough to obtain the accuracy to resolve the
rapidly varying density properties of the medium, a huge amount of
storage and CPU-power is needed, even for the tenuous regions where
less spatial coverage is needed.

In the previous section, we pointed out that our method solves the
radiative transfer equation for every point distribution using the
same simple Markovian random walk process. In this section, we will
exploit this fact and we will use a correlated point process representing
a clumpy medium distribution to demonstrate the optimal resolution
and speed efficiency of our method.

\subsection{Fractal point process}

There are many ways to define a correlated point process, but for
this paper we use a fractal point process to generate the clumpy point
distribution. There are a number of reasons for this choice, the most
important ones being that the process has a simple mathematical definition
and a straightforward implementation, and that its statistical properties
are scale-free by definition, which makes the conclusions we will
draw about the statistics of the resultant tessellation more general.

Fractal (stochastic) point processes have been widely used as a modelling
tool \citep[for a review, see][]{Lowen}. In particular, 
\citet{Mandelbrot} used it to model a non-standard random walk resulting in a
linear L\'{e}vy dust which, he showed, could be used to model
galaxy clusters. Because of this feature, we will use a modified version
of his recipe for constructing the correlated point process, which we will now describe.

A L\'{e}vy flight is a sequence of flights separated by stopovers.
It is constructed by choosing the first stopover randomly and starting
the flight from that point. The (straight-line) flights have the following
properties: their direction is random and isotropic, the different
flights are statistically independent (thus, the L\'{e}vy flight
is a Markov process), and their lengths follow a probability distribution
\begin{equation}
\label{prob_fractal}
f(r)=kr^{d-1-D}=kr^{1-D},
\end{equation}
 where \( d \) is the dimension of the space in which the flights
occur (in our case, \( d=2 \)), \( k \) is a normalisation constant,
and \( D \) is the fractal dimension as defined in \citet{Mandelbrot}.
Eq.(\ref{prob_fractal}) is a modification of the distribution function
used in \citet{Mandelbrot} to the effect that the clustering of points
around \( r=0 \) is avoided.
Thus, if \( D=0 \), we obtain the regular Poisson point process, and if
\( D>0 \), we have an exponential decaying distribution function,
which is scale-free as should be expected from a fractal distribution.
Clustering will increase, when \( D \) is increased. It is only the
stopovers we are interested in, because they will be the points of
the resultant fractal point process. The process is scale-free and
Markovian, by which it is allowed to rescale and translate the resultant
point distribution, so as to center and fit it in our \( [0.0:1.0]^2 \)
domain, without altering the statistical properties.

Of course, we have to choose a lower and upper bound (\( A \)
and \( B \), respectively) for \( r \). Thus, we obtain for the
flight-lengths the cumulative distribution function
\begin{equation}
\label{cum_fractal}
F(r)=\frac{r^{2-D}-A^{2-D}}{B^{2-D}-A^{2-D}}.
\end{equation}
 We fix the upper bound \( B \) (the maximum flight length) as
half the width our domain and choose the lower boundary \( A \) to
be two orders of magnitude smaller. Thus, 
\begin{equation}
\label{Upper_lower_boundaries}
r\in \left[ 0.01,0.5\right].
\end{equation}

\begin{figure}
{\begin{center} \resizebox*{1\columnwidth}{!}{\rotatebox{90}{\includegraphics{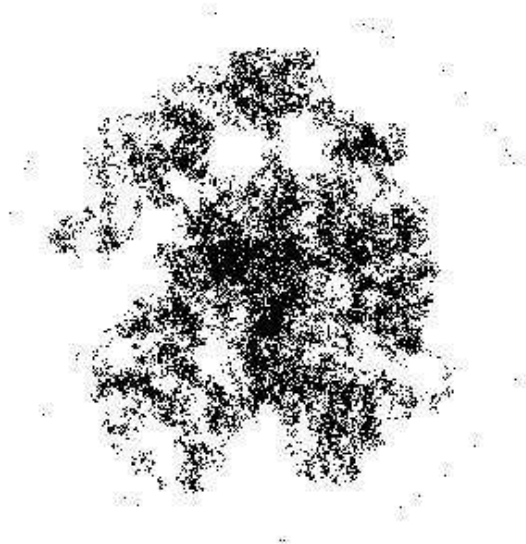}}} \par\end{center}}

\caption{\label{Fractal_Point_proc_fig}Fractal point process satisfying
(\ref{cum_fractal}) and (\ref{Upper_lower_boundaries}) with fractal
dimension \( D=0.5 \) for \( N=10^{5} \)
points. The dots around the fractal distribution are boundary points,
obtained by randomly placing 100 points on the circumference of a
circle with radius 1.1}
\end{figure}

An example of this process satisfying Eq.(\ref{cum_fractal}) and
(\ref{Upper_lower_boundaries}), with \( D=0.5 \) and \( N=10^{5} \),
is shown in Fig.\ref{Fractal_Point_proc_fig}. We will use this point
distribution to illustrate our transfer method in Section 5.4.

\subsection{Length-angle correlation}

The point distribution, which is a result of the fractal point process as 
defined in the previous, gives us the opportunity to assert our claim in
Subsection 3.3 that the angle between two long Delaunay lines is, on average,
smaller than the one between two short lines, by which the angular resolution
is higher for radiation being emitted into optically thin media, which is just
what is needed.
Given a point distribution, it is almost always impossible to find an exact
distribution function \( g(\lambda_{\rmn{D}},\theta ) \) for the correlation between the
Delaunay edge length \(\lambda_{\rmn{D}}\) and the corresponding angle \(\theta\), 
so in order to find the expectation
value for the amount of deflection given a certain length of the Delaunay
line, we do a simple Monte Carlo experiment using a fractal point
process, the result of which is quite general, because the fractal point
process is scale-free.

We construct a point distribution by using the same recipe as used
for making the result in Fig.\ref{Fractal_Point_proc_fig}, but now
we use \( N=2\cdot 10^{5} \) points and define the periodic boundary
conditions \( x=x-\left\lfloor x\right\rfloor  \) and \( y=y-\left\lfloor y\right\rfloor  \).
The periodic boundary conditions will result in a distortion of the
statistical properties of the resultant tessellation, because of the
overlapping parts of the L\'evy flight, but the overall statistical
behaviour on small scales will still have the fractal properties.

\begin{figure}
{\begin{center} \resizebox*{1\columnwidth}{!}{\rotatebox{-90}{\includegraphics{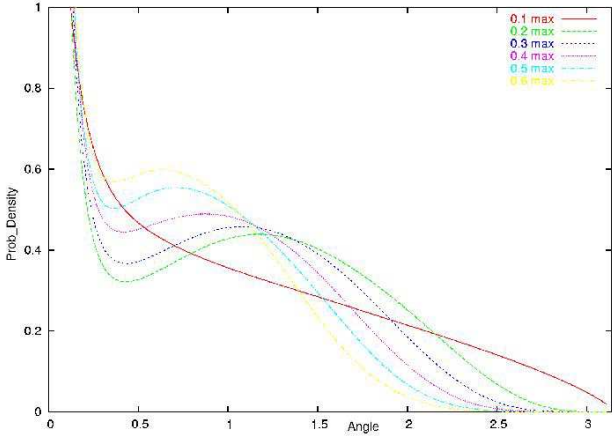}}} \par\end{center}}

\caption{\label{Prob_curves_bins}Plot of the (normalised) probability
functions for the angle between the two most straightforwards paths
for the bins of lengths 0.1, 0.2, ... ,0.6 times the maximum length.}
\end{figure}

At each grid point, we order the connected Delaunay lines of the resultant triangulation
clockwise. Evaluating the average length between two neighbouring lines
of these two lines as well as the angle between them, we can make
a statistic of the length-angle correlation. We sample the lengths
by using \( 20 \) bins going from length \( 0 \) up to the maximum
edge-length (within the triangulation) and the angles by using \( 50 \)
bins in the range \( \left[ 0,2\pi \right]  \). In this way we can
plot a (normalised) distribution function for the angle for each edge-length
bin. The result for several length-bins can be seen in Fig.\ref{Prob_curves_bins}.
The lines were made using B\'{e}zier curves \citep[see][]{Bartels} so as to approximate
the trend of the data-points. Using these B\'{e}zier curves is justified,
because we are only concerned with the trend, or overall behaviour,
and not with very accurate quantitative results. We should remark,
that the distribution functions do go to zero as is required, we would
just have to put more angle-bins near zero. It is interesting to note
the high values of the distribution functions near zero. With a normal
Poisson distribution, the distribution function is \( f(\theta )=\frac{4\sin \theta }{3\pi }\left( (\pi -\theta )\cos \theta +\sin \theta \right)  \)
\citep[e.g][]{Weygaert}, which decreases monotonically when approaching
\( \theta =0 \), so, apparently, the introduction of a (fractal)
correlated point process introduces a large amount of small angles.

\begin{figure}
{\begin{center} \resizebox*{1\columnwidth}{!}{\rotatebox{-90}{\includegraphics{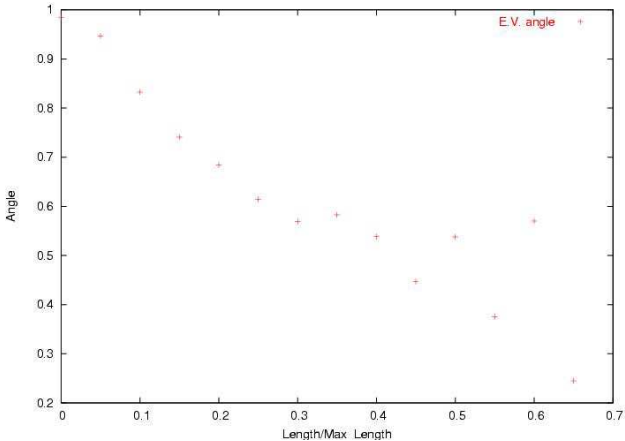}}} \par\end{center}}

\caption{\label{EVAngle_length}Plot of the expectation value of the
angle (in radians) versus the (average) length of the lines.}
\end{figure}

One can readily see that, when the average edge-length is increased,
the average angle between the two lines will, on average, be smaller.
This is noticeable when plotting the expectation value of the angle
versus the length in Fig.\ref{EVAngle_length}. There is a clear downward
trend, with more scatter towards higher lengths because of the statistical
noise. We simply have more data-points at shorter lengths.

Thus, we can conclude that the angle between two long Delaunay lines
(which both originate at the same grid point) will be smaller if the average
edge-length is longer. This is a highly advantageous property of the
Delaunay tessellation based on a correlated point process. Radiation
propagating outwards from a dense region into a tenuous medium will
move onto longer Delaunay lines which connect dense regions. This
means that the angular resolution will be high, as it should be in those regions.

\subsection{Application}

We will now show the results of the application of our radiative transfer
method used on a grid based on a correlated, fractal, point process.
We use the same point distribution as in Fig.\ref{Fractal_Point_proc_fig},
and we put a point source, radiating statically, in the middle of
our \( [0.0:1.0]^2 \) domain. We assign a constant amount of absorption
\( c^{\rmn{abs}} \) to each vertex, and to unambiguously define an absorbing
border we randomly (uniform distribution) place \( 100 \) extra points
on the circumference of a circle with radius \( r=1.1 \) (see Fig.\ref{Fractal_Point_proc_fig}).

\begin{figure*}
\includegraphics[width=17cm]{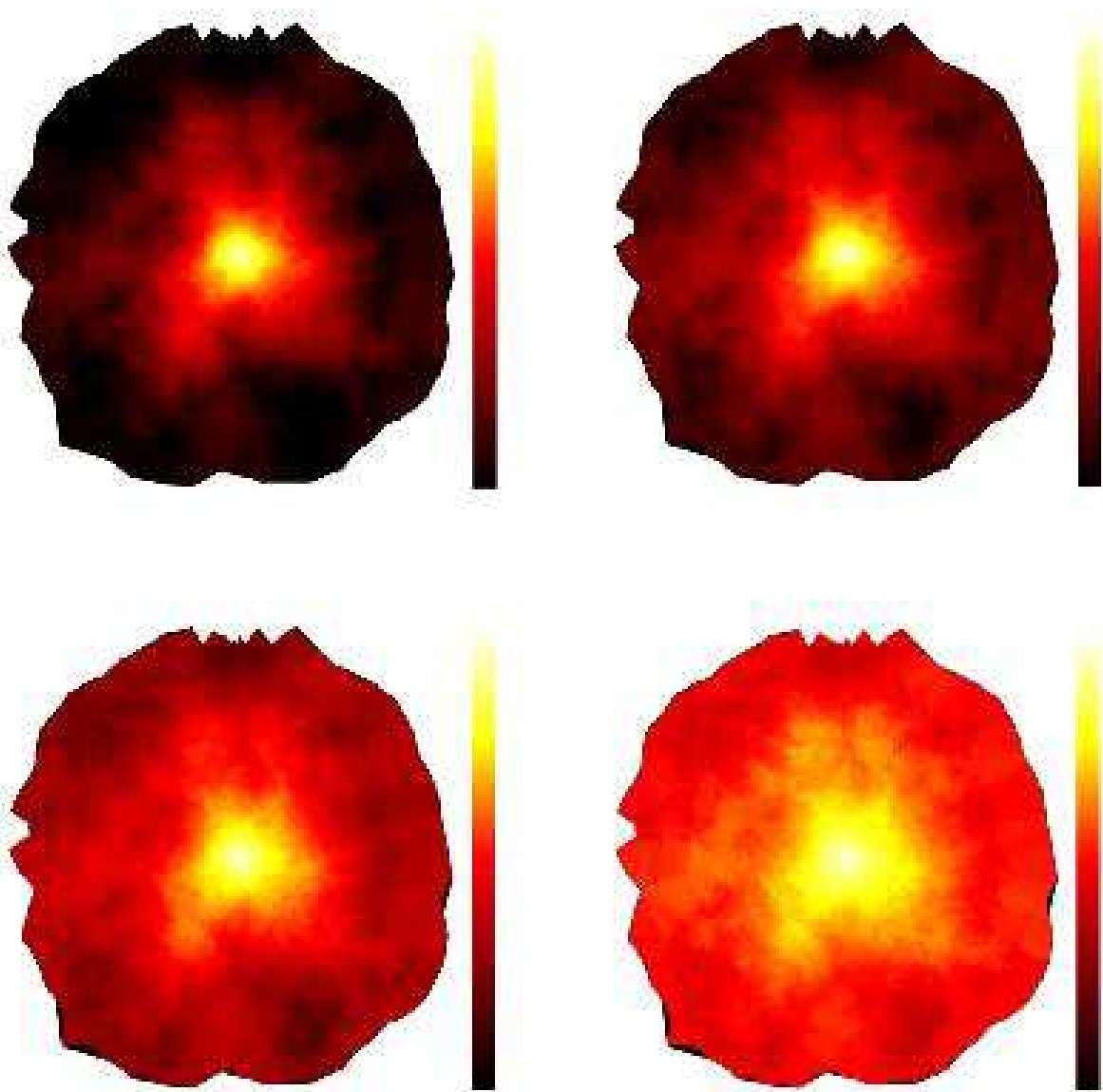}
\caption{\label{Correlated_Results_Fig}Plot of the logarithm of the intensity
of the converged results for (\emph{top}) \( c^{\rmn{abs}}=0.0500 \)
and \( c^{\rmn{abs}}=0.0375 \), and (\emph{bottom})
\( c^{\rmn{abs}}=0.0250 \) and \( c^{\rmn{abs}}=0.0125 \).
Note the shadows behind the dense clumps, especially in the bottom left
image.}
\end{figure*}

The converged results for \( c^{\rmn{abs}}=0.0500 \), \( 0.0375 \), \( 0.0250 \)
and \( 0.0125 \) (we need small \( c^{\rmn{abs}} \)'s, because there is a huge
amount of points) are plotted in Fig.\ref{Correlated_Results_Fig}.
One should note that the intensity is plotted on a logarithmic scale,
by which it is scale-free.

Comparing Fig.\ref{Correlated_Results_Fig} with the point distribution
in Fig.\ref{Fractal_Point_proc_fig}, one immediately sees that the
method works beautifully. All the features of the point distribution
stand out clearly in the radiation results. One readily points out
the high and low density regions, and because of the high resolution
(\( N=10^{5}) \), the distinct features of the high density regions
are resolved very accurately. The exact simulation of shadowing effects
is exemplified, when increasing the absorption coefficient \( c^{\rmn{abs}} \).
With \( c^{\rmn{abs}}=0.0500 \), we clearly see the radiation escaping in the
directions of the lowest point density.

\begin{figure*}
\begin{center}
  \includegraphics[width=17cm]{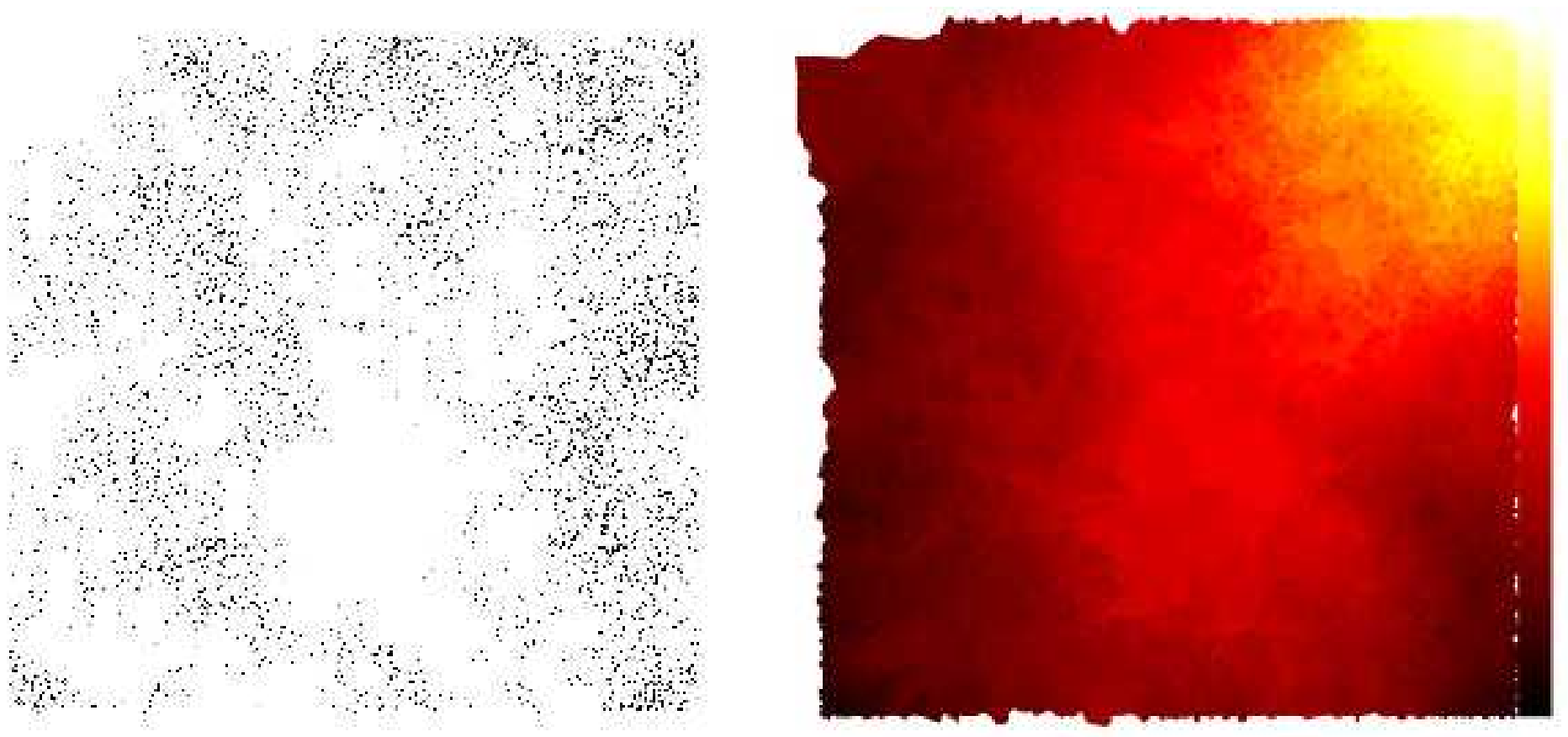} \end{center}
  \caption{Magnified portion \( x,y\in \left[ 0.25,0.50\right] \)
of Fig.\ref{Fractal_Point_proc_fig} (\emph{left}) and Fig.\ref{Correlated_Results_Fig}
(\emph{right}) with \( c^{\rmn{abs}}=0.0500 \).}
\label{Zoom_Fig}
\end{figure*}

The extremely high resolution is pointed out more efficiently, when
zooming in on a part of the domain. Defining the left-bottom corner
of the \( c^{\rmn{abs}}=0.0500 \) result in Fig.\ref{Correlated_Results_Fig}
as \( (0,0) \) and the top-right one as \( (1,1) \), we zoom in
on the region \( x,y\in \left[ 0.25,0.50\right]  \) and plot the
point distribution and radiation result in Fig.\ref{Zoom_Fig}. Notice
the large size of the Voronoi cells filling up the voids in the point
distribution. Even now, the resolution in the dense region is still
high enough to obtain a high amount of accuracy.

\section{Completing the method}

We have shown that our method solves the radiative transfer equations without much
numerical effort, even in those cases in which the medium is highly inhomogeneous.
It is in these cases that our method stands out from other radiative transfer methods,
which have difficulties when passing from one opacity regime to another.

There is one complication, however, as we pointed out earlier in Section 3, because
we treat each event centre as a scattering centre.
When our simulation domain contains large regions of (almost) transparent medium, i.e. when
\(\lambda^{\rmn{scat}}\) becomes bigger than or comparable to the dimensions of our domain,
we can immediately see that this results in an undersampling. 
An extreme example is that of an empty (optically inactive) domain, 
in which our method dictates that no points should be used, by which we do not have a 
grid along which the radiation can propagate. A less extreme example is that of a region
of almost empty (and, thus, undersampled) space behind a highly absorbing clump of matter, 
in which we would like to resolve the sharp shadow cast by the clump. In both cases
the sought-after results are just straight-line trajectories, which are the
solutions of the transfer equations for radiation propagating through a vacuum.

We will now present a supplement to our method, which will enable us to solve the
radiative transfer equations in these cases on the same kind of unstructured grid
based on a Delaunay tessellation.

\subsection{Long characteristics}

Let us focus on the example of a simulation domain which is totally devoid of optically
active medium. Because we need for a grid solving the equations, we proceed by creating
a point distribution. Because the matter is distributed homogeneously across the domain
(it is homogeneously empty), we choose a Poisson point process to generate our point
distribution. The amount of points we choose may vary from case to case and will depend
on the total amount of points available.

Because the solution of the transfer equation in this regime is a superposition of
straight line trajectories and because now each grid point does not represent a packet
of matter at which photons get scattered, we need to modify our photon propagation scheme
as described in Section 3, in such a way that it works according to the long
characteristics principle, as laid out in Subsection 2.2. In order to accommodate this
requirement, we reject the Markov assumption as introduced in Section 3, and we do not
scatter the radiation at each grid point, but we keep track of its original direction,
which is now a very relevant quantity.

\begin{figure}
\begin{center}
\includegraphics[width=4.2cm,clip=]{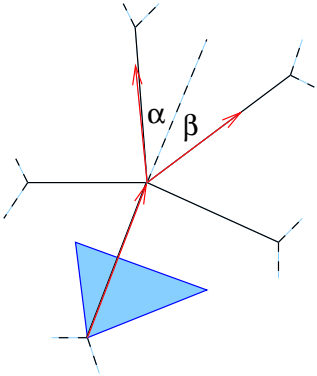}

\caption{\label{LongCharSplit}The advection scheme, as originally laid out in Section 3,
is modified in such a way that at each grid point the radiation is split and redistributed
amongst the \(d\) Delaunay lines which are most `straightforward' with respect to the original
direction (blue dashed line) of the radiation packet.}

\end{center}
\end{figure}

Therefore, we make the modification to the transfer scheme as laid out in Section 3 that
at each intersection we choose the \( d \) most straightforward paths with
respect to the \textit{original} direction (see Fig.\ref{LongCharSplit}). We discuss later
why we choose to split up the radiation into \(d\) parts.

\subsubsection{Mathematical analysis}

\begin{figure}
\begin{center}
{\includegraphics[width=8.4cm,clip=]{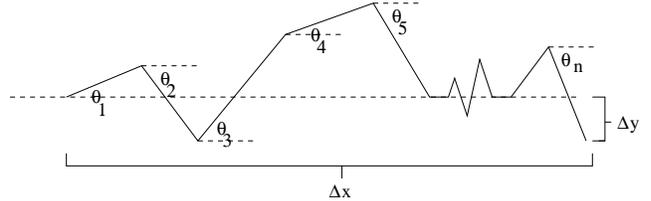}}

\caption{\small \label{Path_figLong}One possible path of a radiation packet performing a walk
of \(n\) steps on the Delaunay graph. The \(i\)-th step is parametrised by an angle \(\theta _i\),
with respect to the original direction \(x\).}
\end{center}
\end{figure}

For a mathematical analysis of the expectation values of the position of a photon packet in
this modified scheme, we proceed as follows. An
example of a path of a photon packet performing a walk in two dimensions is given in Fig.\ref{Path_figLong}.
The following analysis, however, will be valid in \( d \)-dimensional space.

Because of cylindrical symmetry around the original direction \( \bmath{x} \), we can parametrise the
\( i \)-th step by only one angle \( \theta _{i} \), which is the angle between the \( i \)-th
Delaunay edge and the original direction.
Thus, the expectation value of the total displacement
\( \bmath{R}_{n}=\bmath{r}_{1}+...+\bmath{r}_{n} \) is
\begin{eqnarray}
\label{vec_average_long}
\left\langle \bmath{R}_{n}\right\rangle & = & \left\langle \bmath{r}_{1}\right\rangle +...+\left\langle \bmath{r}_{n}\right\rangle \nonumber \\
& = & {n}\lambda_{\rmn{D}}\left\langle \cos \theta \right\rangle \frac{\bmath{x}}{\left| \bmath{x}\right| } \nonumber \\
& = & {n}\lambda_{\rmn{D}}\chi \frac{\bmath{x}}{\left| \bmath{x}\right| },
\end{eqnarray}
in which \( \lambda _{\rmn{D}} \) is defined in Eq.(\ref{AverageDelaunay}), and
\begin{equation}
\label{chi_defined}
\chi = \int ^{\pi} _{-\pi} h(\theta)\cos\theta \rmn{d}\theta.
\end{equation}
\( h(\theta ) \) Is a certain symmetric function, which characterises the probability distribution
of the angle \(\theta\) and which, in most cases, cannot be evaluated analytically.
The second-order expectation value can be evaluated as follows:
\begin{eqnarray}
\label{sec_exp_long}
\left\langle \bmath{R}^{2}_{n}\right\rangle  & = & \left\langle \bmath{r}^{2}_{1}\right\rangle +\left\langle \bmath{r}_{1}\cdot
\bmath{r}_{2}\right\rangle +...+\left\langle \bmath{r}^{2}_{n}\right\rangle \nonumber \\
 & = & \lambda_{\rmn{D}} ^{2}\left({n} + {n}({n-1})\left\langle \cos ( \theta _{i} + \theta _{j} ) \right\rangle\right),
\end{eqnarray}
in which we may choose \( i \) and \( j \) randomly from the set \( \left\{ 1,...,n \right\} \), as long as
\( i \not = j \), because the distribution function \( h(\theta ) \) has the same form for
each angle \( \theta _i \). Using the cosine addition formula, we can reduce Eq.(\ref{sec_exp_long}) to
\begin{equation}
\label{sec_explong_final}
\left\langle \bmath{R}^{2}_{n}\right\rangle = \left({n} + {n}({n-1})\chi^2\right)\lambda_{\rmn{D}} ^{2}.
\end{equation}
Thus, the variance of the displacement is
\begin{equation}
\label{variance_long}
\sigma ^2 _{\bmath{R} _{n}} = {n} \lambda_{\rmn{D}} ^{2}(1-\chi ^2).
\end{equation}
If \( h( \theta ) \propto \delta(\theta) \), then \( \chi = 1 \), by which \( \left\langle \bmath{R} _{n} \right\rangle =
\lambda_{\rmn{D}} {n} \frac{\bmath{x}}{\left| \bmath{x}\right| }\) 
and \( \sigma ^2 _{\bmath{R} _{n}} = 0 \) as should be expected.
The exact form of a distribution function like \( h(\theta) \)
can probably not be evaluated, even in the well-studied Poisson case, but we can use a step function as an approximation. 
Thus, given that in 2D the average number of Delaunay lines meeting at a grid point is \(6\), we use as a step function
\( h(\theta) = 3 / \pi \) on the domain \( \theta \in [-\pi/6,\pi/6] \).
This results in \( \chi = 3/\pi \), by which 
\begin{equation}
\label{exp_long_step}
\left\langle \bmath{R}_{n} \right\rangle = \frac{3n\lambda_{\rmn{D}}}{\pi} \frac{\bmath{x}}{\left| \bmath{x}\right| },
\end{equation}
which is very close (difference of less than \(5\%\))
to the distance along a straight line, which would be \(n\lambda_{\rmn{D}}\). We can always, of course, 
rescale the lengths so as to make sure that the distance traversed equals the exact physical one.
More importantly, the variance in the displacement, in this case, is
\begin{equation}
\label{variance_long_step}
\sigma ^2 _{\bmath{R} _{n}} = \frac{\pi^2 - 9}{\pi^2} \lambda_{\rmn{D}} ^2{n}.
\end{equation}

We know that the results of using a step-function as distribution function gives upper bounds
on the values of Eq.(\ref{vec_average_long}) and Eq.(\ref{variance_long}), because the actual
distribution function would peak around \( \theta = 0 \) and would decrease as \( \left| \theta \right| \) 
increases, so we expect the actual value of \( \sigma ^2 _{\bmath{R} _{n}} \) to be smaller.
Thus, we can simulate a straight line trajectory with this method, because \( \bmath{R} _{n}\propto{n}\bmath{x}\),
but, still, the standard deviation will increase with \( \sqrt{n} \).

What is more important is the behaviour of the standard deviation, if the number of
grid points \( N \) increases. Let us therefore examine a line segment in the simulation
domain of length \( L \) (\(\leq \sqrt{d}\), if we have a \([0.0:1.0]^d\) domain). 
Because the point distribution is homogeneous, we can conclude 
that the number of steps to cover the line is
\begin{equation}
\label{line_steps}
{n}=\xi {N}^{1/d},
\end{equation}
in which \( \xi \leq \frac{\pi}{3}\zeta\sqrt{d} \), which can be found by using the upper bound Eq.(\ref{exp_long_step}) and
the Eq.(\ref{AverageDelaunay}) for the length \(\lambda_{\rmn{D}}\) of a Delaunay line.
If we combine Eq.(\ref{variance_long_step}) with Eq.(\ref{line_steps}), again using Eq.(\ref{AverageDelaunay}), we obtain
\begin{equation}
\label{sigma_long_number}
\sigma \propto \lambda_{\rmn{D}} \sqrt{n} \propto {N}^{-1/2d}.
\end{equation}
Thus, we can conclude that the amount of widening of the beam
will go to zero, if we increase the amount of grid points \(N\).

Even if we do not have a large amount of points to suppress the widening of the beam,
we have another effect which compensates for the widening. Namely, at each intersection
the radiation is split up into \( d \) parts. This means that the intensity at points
farther away from the straight line trajectory is much less than at points close by, simply
because of the fact that more paths cross each other at points close to the line.

\subsubsection{Implementation}

There are several easy algorithms to numerically simulate a straight line across
a Delaunay graph, most having their origin in mobile telecommunications 
\citep[see][]{Bacelli}. These algorithms determine the shortest path along a
Delaunay-graph from one point to another, both of which lie on a line.
So, why not use one of these algorithms, which can be implemented into our method
without much effort, to accurately model a straight line, instead
of using the one we described which introduces a minor widening of the beam?

The main reason is that, when we have a point source, there are only so many rays,
or Delaunay edges, emerging from that point source. This means that parts of the
volume at large distances from the source are largely undersampled, with only a couple
or none of the rays intersecting it. This is the main drawback of the usual methods
 which we criticised in Section 2. Therefore, we still use the feature of our
scheme, that we split up the radiation at each intersection into \( d \) parts. This
introduces a \( \sqrt{n} \) widening of the beam, but it also makes sure that the
whole domain will be covered. In this sense, it shares some of the characteristics
of the adaptive ray tracing mechanism as described in \citet{Abel2}, which
describes a way of splitting up rays, so as to accurately sample the whole domain, but,
by doing so, also introduces a similar widening of the beam.

Another reason is that unless the low density regions are extremely big, the region in
which one might want to use this long characteristics method is not very large, so the
effect of beam widening is not of much importance, even if we only have a small number
of points, because the number of steps \( n \) is small.

\begin{figure}
\begin{center}
{\includegraphics[width=8cm,clip=]{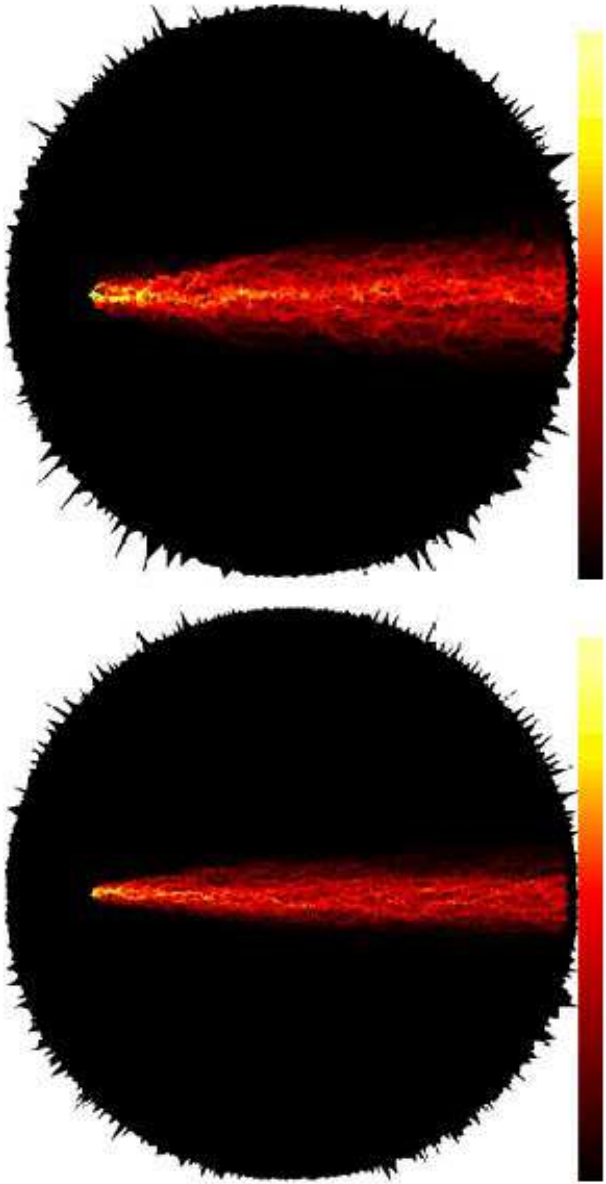}}
%{\begin{center} \resizebox*{1\columnwidth}{!}{\rotatebox{90}{\includegraphics{./laser.ps}}} \par\end{center}}

\caption{\label{laser_fig}Results of using the long characteristics version of our
method in order to simulate a pencil beam, using 30000 (top) and
100,000 points (bottom). The absorption coefficient is equal to zero.}
\end{center}
\end{figure}

To get an idea of the algorithm and how it works, we try to simulate a laser beam in
a transparent medium. We use a Poisson process to create a homogeneous
point distribution of \( 3\times 10^4\) and \( 10 ^5 \) points and place a pencil beam
in the domain. The result can be seen in Fig.\ref{laser_fig}. Note that the absorption
coefficient is equal to zero.
As expected, the beam is narrower when the number of points is larger, 
as should be expected from Eq.(\ref{sigma_long_number}), and the intensity
is highest close to the straight line trajectory.

\begin{figure}
\begin{center}
{\includegraphics[width=8cm,clip=]{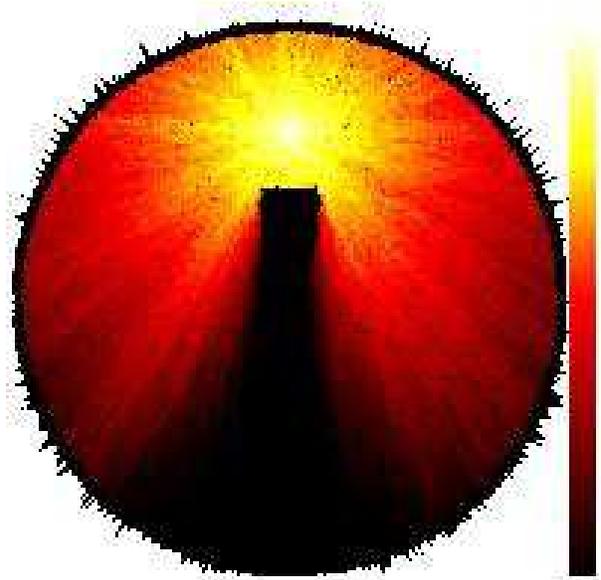}}
%{\begin{center} \resizebox*{1\columnwidth}{!}{{\includegraphics{./Figs/LongShadow.ps}}} \par\end{center}}

\caption{\label{point_long_fig}Radiation from a point source hitting a highly absorbing square, which casts a sharp shadow. Result obtained by using the long characteristics variant with \( 10^5 \) points, homogeneously distributed.}
\end{center}
\end{figure}
These properties make this variant ideal to resolve sharp shadows behind highly
absorbing objects. To illustrate this, we simulate a point source by defining all points
within a circle of radius \( 0.1 \) with \( (0.5,0.8) \) as a centre to be 
point sources, and we put a highly absorbing object in front of our source. The object,
in this case, is a square with side \( 0.1 \) which is centred on \( (0.5,0.65) \), with
\( c^{\rmn{abs}}=1.0 \) at each point. The rest of the domain (except the absorbing boundaries, of
course) has \( c^{\rmn{abs}}=0.1 \). The result for \(10^5\) points is plotted in Fig.\ref{point_long_fig}.

\subsection{Usage}

As we have shown in the previous subsection, the long characteristic variant of our
method results in numerical solutions in which shadows can be resolved quite accurately.
So, why do we not use this long characteristic variant as our main method to solve
the radiative transfer equation numerically? There are several reasons.

First, the long characteristic variant is only useful when there are large, almost empty
regions, in which there would be no scattering. Because the biggest difficulty lies in
solving the equation for photons propagating through media of various densities, we
have made sure that the core of our method solves this difficult problem.  The second issue is speed. 
The improved angular resolution of this variant
does not come without a cost. For \( i \) point sources, the method will in 2D
be \( 6i \) (on average \( 6 \) Delaunay-edges per nucleus) times slower 
than with the original method, which has an operation count of \( O(N) \)
independent of the number of sources.

Thus, we have presented two methods \citep[excluding the long characteristics variant from][]{Bacelli}
which are able to solve the transfer equations on unstructured grids, such as a Delaunay
graph. Each has its own advantages and disadvantages and it is basically up to the user 
to choose which version is most suited for the problem at hand.

For now, we will choose to use the first version of our method as a basis,
making use of its speed, efficiency and adaptability to solve the transfer equation, when
it is most difficult to solve, and we shall only resort to the slower long characteristics variant
in those cases where there are large transparent regions, or in which high resolution
shadows are needed.

\section{Conclusions and future work}

We have presented a new numerical method that is able to solve the
transfer equation efficiently on unstructured grids, such as a Delaunay
graph, which are based on random point processes. One of its main
advantages is that it uses a Lagrangian grid, which puts the accuracy
where it is needed, and it automatically puts, as we have shown, the angular resolution 
where it is needed. We have shown that if we choose a point distribution
in the from of Eq.(\ref{PointDensToDens}) to mimic the medium density profile,
we obtain a set of global constants, or interaction coefficients \(\{c^i\}\),
which are assigned to each grid point, or event centre. This procedure ensures
that, when the radiation performs a Markovian random walk along the graph 
from one event centre to the next, the overall macroscopic behaviour of the 
radiation field is just as we expect from the radiative transfer equations.
One can intuitively understand this, because, if we assign the same amount of 
withheld radiation \( c^i \) to each grid point and we put more points where the medium is
more dense, that region will automatically become more opaque. 
Moreover, because the point distribution adapts to the medium distribution
and because the algorithm makes no distinction between optically thick and optically
thin regions, our new method can be used equally as well in every opacity regime, which
makes this method particularly suitable to be used in those realistic cases, where
the medium passes from one regime to the next. It is in these cases that most
other methods fall short. But the most important advantage of all
is that we have reduced the complex system of coupled differential
equations to a simple one-dimensional random walk on a graph, in which
the interaction recipes in the form of Eqs.(\ref{AbsRecipeAbs}) and (\ref{AbsRecipeOut}) are the most difficult calculation to be performed. Therefore,
the method solves the transfer equation in \( O\left(N^{1+1/d}\right) \) operations,
in which \( N \) is the number of resolution elements, or grid points,
even in the cases where the number of sources approaches \( N \),
for which the operation count of other \( O(N) \) schemes \citep[e.g.][]{Abel1} 
increases towards higher orders. This makes our method extremely suitable
for use in cases with large extended sources distributed over space. Any such implementation 
will therefore be extremely fast. 

We have also described a supplement to our method, which can be used, if
the domain contains large almost empty regions, by which it would be highly
undersampled. This long characteristics variant can be used to accurately
resolve shadows behind highly absorbing objects, but it comes with a cost.
When the number of sources is increased to \( N \), the operation count
will increase to \( O(N^2) \).

There are several things that remain to be done and on which we are
working already. First, we are extending our method to three dimensions.
This is no problem, because our code is set up generically and
the mathematical analysis is valid for \( d \)-dimensional space.
Another thing is that we have not incorporated
feedback from the medium at all. We want to apply our method to cases
in which the medium is optically active and the source function is
inhomogeneous and frequency-dependent, or which incorporates
ionisation, recombination and photodissociation. That is, we want to incorporate
a wider variety of interaction coefficients \( \{c^i\}\). A most promising feature
of our method is that it also inherently solves the time-\emph{dependent}
radiative transfer equations without doing any extra work. Because
our method is photon-conserving, we can use the time-dependent variant,
for example, to accurately model ionisation fronts expanding in an
inhomogeneous medium. Finally, it is our aim to combine our method with the SPH
method, or even other hydrodynamical schemes, so that we can combine
the two essential parts of the physics which probably governs the
formation of most structures in our universe.

\section*{Acknowledgments}

\label{lastpage}


\begin{thebibliography}{99}
\bibitem[\protect\citeauthoryear{Abel et al.}{1999}]{Abel1}Abel, T., Norman, M. L. \& Madau, P. 1999, ApJ, {\bf 523}, 66
\bibitem[\protect\citeauthoryear{Abel \& Wandelt}{2002}]{Abel2}Abel, T. \& Wandelt, B. D. 2002, MNRAS, {\bf 330}, L53-L56
\bibitem[\protect\citeauthoryear{Baccelli et al.}{1998}]{Bacelli}Baccelli, F., Tchoumatchenko, K. \& Zuyev, S. 1998, {\it Markov Paths on the Poisson-Delaunay Graph}, INRIA Rapports de Recherche, Nr. 3420
\bibitem[\protect\citeauthoryear{Barber et al.}{1996}]{Barber}Barber, C. B., Dobkin, D. P. \& Huhdanpaa H. 1996, {\it ACM Transactions on Mathematical Software}, {\bf 22}, No. 4, 469-483
\bibitem[\protect\citeauthoryear{Bartels et al.}{1998}]{Bartels}Bartels, R. H, Beatty, J. C. \& Barsky, B. A. 1998, {\it B\'ezier Curves},  Ch. 10 in {\it An Introduction to Splines for Use in Computer Graphics and Geometric Modeling},
San Francisco, CA: Morgan Kaufmann, pp. 211-245, 1998
\bibitem[\protect\citeauthoryear{Chandrasekhar}{1943}]{Chandrasekhar}Chandrasekhar, S. 1943, Rev. Modern Phys., {\bf15}, 1-87
\bibitem[\protect\citeauthoryear{Delaunay}{1934}]{Delaunay}Delaunay, B. 1934, Classe des Sciences Math\'ematiques et Naturelles, {\bf7}, 793
\bibitem[\protect\citeauthoryear{Duderstadt \& Martin}{1979}]{Duderstadt}Duderstadt, J. J. \& Martin W. R. 1979, {\it Transport Theory},
John Wiley and Sons, New York 
\bibitem[\protect\citeauthoryear{Greengard}{1988}]{Greengard}Greengard, L. F. 1988, {\it The Rapid Evolution of Potential Fields in Particle Systems}, MIT Press 
\bibitem[\protect\citeauthoryear{Kunasz \& Auer}{1988}]{Kunasz}Kunasz, P. B. \& Auer, L. 1988, JQSRT, {\bf 39}, 67
\bibitem[\protect\citeauthoryear{LeVeque}{1998}]{LeVeque}LeVeque, R. J., Mihalas, D., Dorfi, E. A., et al.
1998, pp. 104-111 in: {\it Computational Methods for Astrophysical Fluid Flow}, Saas-Fee Advanced Course 27, eds. O. Steiner \& A.
Gautschy, Springer Verlag, Berlin
\bibitem[\protect\citeauthoryear{Lowen \& Teich}{1995}]{Lowen}Lowen, S. B. \& Teich, M. C. 1995, {\it Fractals}, {\bf 3}, No. 1 
\bibitem[\protect\citeauthoryear{Lucy}{1977}]{Lucy}Lucy, L. B. 1977, AJ, {\bf 82}, 1013 
\bibitem[\protect\citeauthoryear{Mandelbrot}{1982}]{Mandelbrot}Mandelbrot, B. B. 1982, {\it The Fractal Geometry of Nature}, W.
H. Freeman \& Co. 
\bibitem[\protect\citeauthoryear{Mihalas et al.}{1978}]{Mihalas}Mihalas, D., Auer, L. \& Mihalas, B. 1978, ApJ, {\bf 220}, 1001
\bibitem[\protect\citeauthoryear{Monaghan}{1992}]{Monaghan}Monaghan, J. J. 1992, ARA\&A, {\bf 30}, 543 
\bibitem[\protect\citeauthoryear{Okabe et al.}{2000}]{Okabe}Okabe, A., Boots, B., Sugihara, K., \& Chiu, S.N. 2000, {\it Spatial Tessellations: Concepts and Applications of Voronoi Diagrams}, 2nd
Ed., Chichester, John Wiley \& Sons Ltd 
\bibitem[\protect\citeauthoryear{Pelupessy et al.}{2003}]{Pelupessy}Pelupessy, F. I., Schaap W. E. \& Van de Weygaert, R. 2003, A\&A,
{\bf 403}, 389
\bibitem[\protect\citeauthoryear{Rutten}{1999}]{Rutten}Rutten, R. 1999, {\it Radiative Transfer in Stellar Atmospheres}, 
\texttt{http://www.fys.ruu.nl/\textasciitilde{}rutten/}
\bibitem[\protect\citeauthoryear{Schaap \& Van de Weygaert}{2000}]{Schaap}Schaap, W. E., \& Van de Weygaert, R. 2000, A\&A, {\bf 363}, L29 
\bibitem[\protect\citeauthoryear{Steinacker et al.}{2002}]{Steinacker}Steinacker, J., Hackert, R. \& Steinacker, A. 2002,
JQSRT, {\bf 74}, 183
\bibitem[\protect\citeauthoryear{Stoyan et al.}{1996}]{Stoyan}Stoyan, D., Kendall, W. S. \& Mecke, J. 1996, {\it Stochastic Geometry and its Applications}, 2nd Ed., John Wiley \& Sons 
\bibitem[\protect\citeauthoryear{Van de Weygaert}{1991}]{Weygaert}Van de Weygaert, R., {\it Voids and the Geometry of Large Scale Structure}, Ph.D. Thesis, Leiden 1991 
\bibitem[\protect\citeauthoryear{Voronoi}{1908}]{Voronoi}Voronoi, G. 1908, Journal f\"ur die Reine und Angewandte Mathematik, {\bf134}, 198
\end{thebibliography}
\end{document}